\documentclass[%
 reprint,
superscriptaddress,
 amsmath,amssymb,
 aps,
]{revtex4-2}

\usepackage{graphicx}
\usepackage{dcolumn}
\usepackage{bm}
\usepackage{hyperref}
\usepackage{mathtools}

\usepackage{physics}
\usepackage{feynmp}
\usepackage{amssymb}
\usepackage{amsmath}
\usepackage{color}
\usepackage{braket}
\usepackage{array}
\usepackage{ulem}

\usepackage{booktabs}
\usepackage{tabularx}
\usepackage{multirow}
\usepackage{array}
\newcolumntype{Y}{>{\centering\arraybackslash}X}
\newcolumntype{Z}{>{\raggedright\arraybackslash}X}
\newcolumntype{L}{>{\raggedright\arraybackslash}p{1cm}}
\newcolumntype{W}{>{\centering\arraybackslash}p{2cm}}

\allowdisplaybreaks[4]

\begin{document}

\preprint{APS/123-QED}

\title{$\Xi NN$ three-baryon force from SU(3) chiral effective-field theory: A femtoscopic study}

\author{Gen Uratsu}
 \email{uratsu.gen.456@s.kyushu-u.ac.jp}
 \affiliation{Department of Physics, Kyushu University, Fukuoka 819-0395, Japan.}

\author{Tokuro Fukui}%
 \email{tokuro.fukui@artsci.kyushu-u.ac.jp}
 \affiliation{Faculty of Arts and Science, Kyushu University, Fukuoka 819-0395, Japan.}
 \affiliation{RIKEN Nishina Center, Wako, 351-0198, Japan}

\author{Kazuyuki Ogata}
 \email{ogata.kazuyuki.169@m.kyushu-u.ac.jp}
 \affiliation{Department of Physics, Kyushu University, Fukuoka 819-0395, Japan.}

\date{\today}

\begin{abstract}
 \edef\oldrightskip{\the\rightskip}
\begin{description}
 \rightskip\oldrightskip\relax
 \item[Background]
	    The development of SU(3) chiral effective field theory has opened the way to a systematic exploration of three-baryon forces (3BFs), a key ingredient in hypernuclear and dense matter physics.
        However, $\Xi NN$ 3BF based on SU(3) chiral EFT has not been studied until now.
 \item[Purpose]
	    We apply SU(3) chiral EFT to derive $\Xi NN$ potentials in momentum space.
        Then, we investigate how the $\Xi NN$ 3BF affects the correlation function of deuteron--$\Xi^-$ pair created through heavy-ion collisions.
 \item[Methods]
	    To reduce the number of low-energy constants involved in the $\Xi NN$ potentials, we employ the decuplet saturation approximation, by which only two of them remain unconstrained.
        The deuteron–$\Xi^-$ scattering is treated as an effective two-body problem with the $\Xi NN$ 3BF incorporated into the potential between the deuteron and $\Xi^-$.
 \item[Results] 
	    We found that the effect of the $\Xi NN$ 3BF on the deuteron--$\Xi^-$ correlation function is at most about 4\%.
        This small effect is not primarily due to the loosely-bound nature of the deuteron.
        Instead, this is because the deuteron and $\Xi^-$ interact with each other mainly at low momentum, corresponding to peripheral scattering, where the influence of the $\Xi NN$ 3BF is limited.
 \item[Conclusions]
        Since the correlation function shows limited sensitivity to the short-range 3BF, complementary approaches may be necessary.
\end{description}
\end{abstract}

\maketitle


\section{Introduction}
\label{sec:intro}

Three-baryon forces (3BFs), though not directly measurable and inherently scheme-dependent, are essential for understanding hypernuclear phenomena.
One notable example is the inclusion of a $\Lambda NN$ 3BF, modeled as a two-pion exchange (TPE) plus phenomenological repulsive term, in calculations of hyperneutron matter~\cite{PhysRevLett.114.092301}.
More recently, lattice simulations of hyperneutron matter revealed the effect of $\Lambda NN$ and $\Lambda\Lambda N$ 3BFs, which are defined under the pionless effective-field theory (EFT)~\cite{TONG2025825,Tong_2025}.

To systematically investigate the meson-exchange contributions to 3BFs beyond pionless picture, SU(3) chiral EFT~\cite{10.3389/fphy.2020.00012} serves as a suitable framework, currently formulated up to next-to-next-to-leading order (N$^2$LO)~\cite{HaidenbauerEPJA2023}.
At N$^2$LO in SU(3) chiral EFT, leading 3BFs emerge, and a general framework was proposed for deriving the corresponding potentials~\cite{PhysRevC.93.014001}. 
This framework has been applied to derive $\Lambda NN$ 3BF, the role of which was investigated in light hypernuclei~\cite{Le2025,PhysRevLett.134.072502} and hyperneutron matter~\cite{Gerstung2020}.
See a recent review~\cite{haidenbauer2025abinitiodescriptionhypernuclei} for the \textit{ab initio} description of hypernuclei.
A natural extension would be its application to the strangeness $S = -2$ sector, although no such studies have been carried out to date.

To investigate 3BFs with $S=-2$, $\Xi NN$ systems serve as promising candidates. 
So far, various aspects of the $\Xi NN$ systems, such as bound states~\cite{PhysRevLett.110.012503,Garcilazo_2014,Garcilazo_2015,PhysRevC.93.034001,Filikhin2017,Miyagawa2021,Egorov2023}, resonant states~\cite{PhysRevC.93.024001,Garcilazo_2020}, and scattering~\cite{PhysRevC.57.2858}, have been predicted based on the Faddeev approaches~\cite{MIYAGAWA1997535,Gloeckle2001}.
In addition, the Gaussian expansion method~\cite{HIYAMA2003223} has been employed for studying the $\Xi NN$ bound states~\cite{PhysRevLett.124.092501}.
However, none of these studies take into account the $\Xi NN$ 3BF.

Recently, a femtoscopic analysis of the deuteron--$\Xi^-$ correlation function was performed in Ref.~\cite{PhysRevC.103.065205}, where direct deuteron production was found to dominate over production via final-state interaction in heavy-ion collisions,
although the $\Xi NN$ 3BF was not considered there.
In comparison, the effects of three-nucleon forces on femtoscopic observables was suggested to be small in the proton--deuteron case~\cite{PhysRevX.14.031051}, where the Urbana IX three-nucleon force~\cite{PhysRevC.56.1720} was used in combination with the Argonne $V18$ two-nucleon force~\cite{PhysRevC.51.38}.
However, the situation may differ for hyperonic 3BFs. 
In particular, a recent study~\cite{PhysRevC.110.054004} showed that phenomenological $\Lambda NN$ 3BFs significantly reduce the $pp\Lambda$ correlation function. 
Although such analyses are model-dependent, they clearly demonstrate the possible impact of hyperonic 3BFs. 
This motivates us to examine whether similar effects arise from the $\Xi NN$ 3BF in the deuteron–$\Xi^-$ correlation function.

In this study, we derive the $\Xi NN$ 3BF based on SU(3) chiral EFT.
To assess its physical relevance, we apply the decuplet saturation approximation (DSA)~\cite{PETSCHAUER2017347}, which reduces the number of low-energy constants (LECs) involved in the three-baryon potentials.
We then explore the effect of the $\Xi NN$ 3BF on the deuteron–$\Xi^-$ correlation function by modeling their scattering as an effective two-body problem, where the $\Xi NN$ 3BF is incorporated into the potential between the deuteron and $\Xi^-$.


This paper is organized as follows: 
In Sec.~\ref{sec:form}, we formulate the general expression of the $\Xi NN$ 3BF as potentials in momentum space. 
We also give formulas necessary for implementing the potentials in the calculations of the correlation function.
In Sec.~\ref{sec:res}, we demonstrate how the $\Xi NN$ 3BF affects the correlation function.
Section~\ref{sec:summary} is devoted to the summary and perspectives of this work.

\begin{widetext}
\section{Formalism}
\label{sec:form}
\subsection{Derivation of $\Xi NN$ potential}
\label{sec:form_XiNNpot}
\subsubsection{Overview of derivation}
\label{sec:form_XiNNpot_overview}
\begin{figure}[!t]
    \centering
    \includegraphics[width=0.7\linewidth]{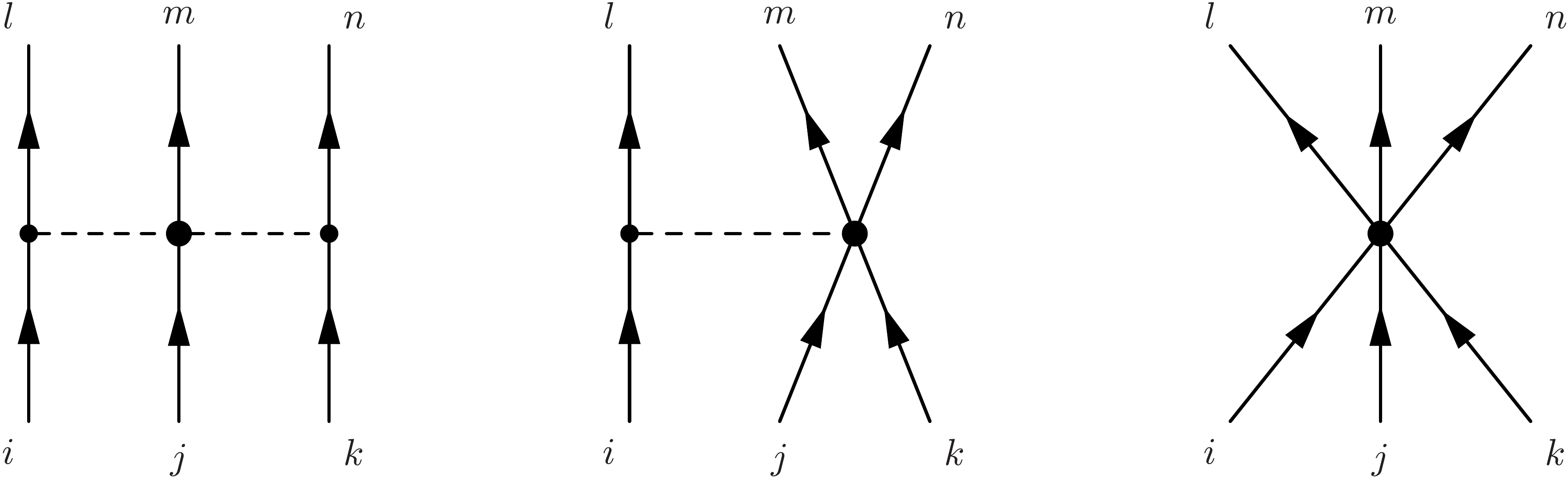}
    \caption{Feynman diagrams for the leading $\Xi NN$ 3BF.
    From left to right, they correspond to the TPE, OPE, and contact terms.
    The arrows specified by the indices $i$, $j$, and $k$ correspond to the initial baryons, while those with $l$, $m$, and $n$ are associated with the final ones.
    The dashed line represents the pion propagation.}
    \label{fig:diagram}
\end{figure}
Based on SU(3) chiral EFT~\cite{PhysRevC.93.014001} we formulate the $\Xi NN$ 3BF,
which consists of the TPE, one-pion exchange plus contact (OPE), and contact terms. 
The basic steps to derive the potentials of the $\Xi NN$ 3BF in momentum space are as follows.
First, the particle-basis potential $v_{ijk,lmn}$ is obtained.
Here, the indices $i$, $j$, and $k$ ($l$, $m$, and $n$) specify the initial (final) baryons, as illustrated in Fig.~\ref{fig:diagram}.

Second, the isospin-basis potential $V_{T_{12}T_{12}'TM_T}^{\Xi NN}$ is calculated through
\begin{align}
    V_{T_{12}T_{12}'TM}^{\Xi NN}
    &=
    \dfrac{1}{2T+1}\sum_{m_i m_j m_k} \sum_{m_l m_m m_n} \sum_{M_{12}M_{12}'} v_{ijk,lmn}
    \notag\\
    &\times \left(\left. \frac{1}{2} m_i \frac{1}{2} m_j \right| T_{12} M_{12} \right) 
    \left(\left. T_{12} M_{12} \frac{1}{2} m_k \right| T M \right)
    \left(\left.\frac{1}{2} m_l \frac{1}{2} m_m \right| T_{12}' M_{12}' \right) 
    \left(\left. T_{12}' M_{12}' \frac{1}{2} m_n \right| T M \right),
    \label{Visospin}
\end{align}
where $T_{12}$ ($T_{12}'$) is the two-particle isospin of the initial (final) state
and $M_{12}$ ($M_{12}'$) is its third component.
The total isospin is denoted by $T$ and its third component is $M_T$.
Because both nucleon and $\Xi$ have isospin $1/2$, the third components $m_i$, $m_j$, $m_k$, $m_l$, $m_m$, and $m_n$ of the isospin can all take the values $\pm1/2$.
The symbol $(\cdot \cdot \cdot \cdot | \cdot \cdot)$ is the Clebsch--Gordan coefficient.

Equation~\eqref{Visospin} is applied to each term, i.e.,
the TPE potential $V_{\mathrm{TPE}}^{\Xi NN}$, 
the OPE potential $V_{\mathrm{OPE}}^{\Xi NN}$,
and contact potential $V_{\mathrm{ct}}^{\Xi NN}$, 
which are all given within the isospin-basis representation.
Note that, throughout the paper, the subscript ``ct'' represents the contact term, while we omit the subscripts specifying the isospin states for simplicity.
Thus, we obtain the total potential as ${V_{\mathrm{TPE}}^{\Xi NN}+V_{\mathrm{OPE}}^{\Xi NN}+V_{\mathrm{ct}}^{\Xi NN}}$.

For later convenience, we define the momentum transfer by 
${\bm{q}_{li} = \bm{p}_l - \bm{p}_i}$ with the initial and final momenta $\bm{p}_i$ and $\bm{p}_l$, respectively.
The potentials to be derived depend on $\bm{q}_{li}$, $\bm{q}_{mj}$, and $\bm{q}_{nk}$,
as well as on spin and isospin operators represented by the Pauli matrices $\bm{\sigma}_i$ and $\bm{\tau}_i$, respectively.
Throughout this paper, we employ natural units with $\hbar = c = 1$.

\subsubsection{Two-pion exchange term}
\label{sec:form_XiNNpot_TPE}
Following the notation of Ref.~\cite{PhysRevC.93.014001}, the two-meson exchange (TME) potential for general three-baryon systems within the particle-basis representation reads
\begin{align}
    v_{\mathrm{TME}}
    &=
    -\dfrac{1}{4f_0^4}\dfrac{\left(\bm{\sigma}_i\cdot\bm{q}_{li}\right) \left(\bm{\sigma}_k\cdot\bm{q}_{nk}\right)}{\left(q_{li}^2+m_{\phi_1}^2\right)\left(q_{nk}^2+m_{\phi_2}^2\right)}
    \left[N_1^{\prime}+N_2^{\prime}\bm{q}_{li}\cdot\bm{q}_{nk}+N_{3}^{\prime}i\bm{\sigma}_j\cdot(\bm{q}_{li}\times\bm{q}_{nk})\right],
    \label{v2meson}\\
    N_1^\prime
    &=
    N_{B_lB_i\bar{\phi}_1}N_{B_nB_k{\phi}_2}\sum_{c^f=b_D,b_F,b_0}\dfrac{c^f}{4}\left(N^f_{\phi_1\substack{m\\j}\bar{\phi}_2}+N^f_{\bar{\phi}_2\substack{m\\j}\phi_1}\right),
    \label{N1_2meson}\\
    N_2^\prime
    &= -N_{B_lB_i\bar{\phi}_1}N_{B_nB_k\phi_2}
    \sum_{c^f=b_1,b_2,b_3,b_4}
    c^f\left(N^f_{\phi_1\substack{m\\j}\bar{\phi}_2}+N^f_{\bar{\phi}_2\substack{m\\j}\phi_1}\right),
    \label{N2_2meson}\\
    N_3^\prime
    &=N_{B_lB_i\bar{\phi}_1}N_{B_nB_k\phi_2}
    \sum_{c^f=d_1,d_2,d_3}
    c^f\left(N^f_{\phi_1\substack{m\\j}\bar{\phi}_2}-N^f_{\bar{\phi}_2\substack{m\\j}\phi_1}\right).
    \label{N3_2meson}
\end{align}
Here, for simplicity, the indices $i$, $j$, $k$, $l$, $m$, and $n$ are omitted from the left-hand side of Eq.~\eqref{v2meson},
in which $f_0$ is the pion decay constant in the chiral limit 
and $m_{\phi_i}$ is the mass of meson $\phi_i$.
The LECs representatively expressed by $c^f$ are associated with the meson--baryon scattering.

The coefficient $N_{BB\phi}$ determines the three-point vertex, as illustrated in Fig.~\ref{fig:2meson}(a), and obtained by comparing two equivalent forms of the chiral effective Lagrangian for meson--baryon couplings~\cite{doi:10.1142/S0218301395000092}; one written in matrix notation and the other in the particle-basis representation~\cite{PhysRevC.93.014001}.
Similarly, $N^f_{\phi\substack{m\\j}\phi}$ associated with the diagram in Fig.~\ref{fig:2meson}(b) can be evaluated using the meson--baryon chiral effective Lagrangian~\cite{Krause1990,JoseAntonioOller_2006} in matrix form and that in the particle-basis representation~\cite{PhysRevC.93.014001}.
The explicit values of these SU(3) coefficients are relegated to Appendix~\ref{sec:vertexcoef}.

\begin{figure}[!t]
    \centering
    \includegraphics[width=0.6\linewidth]{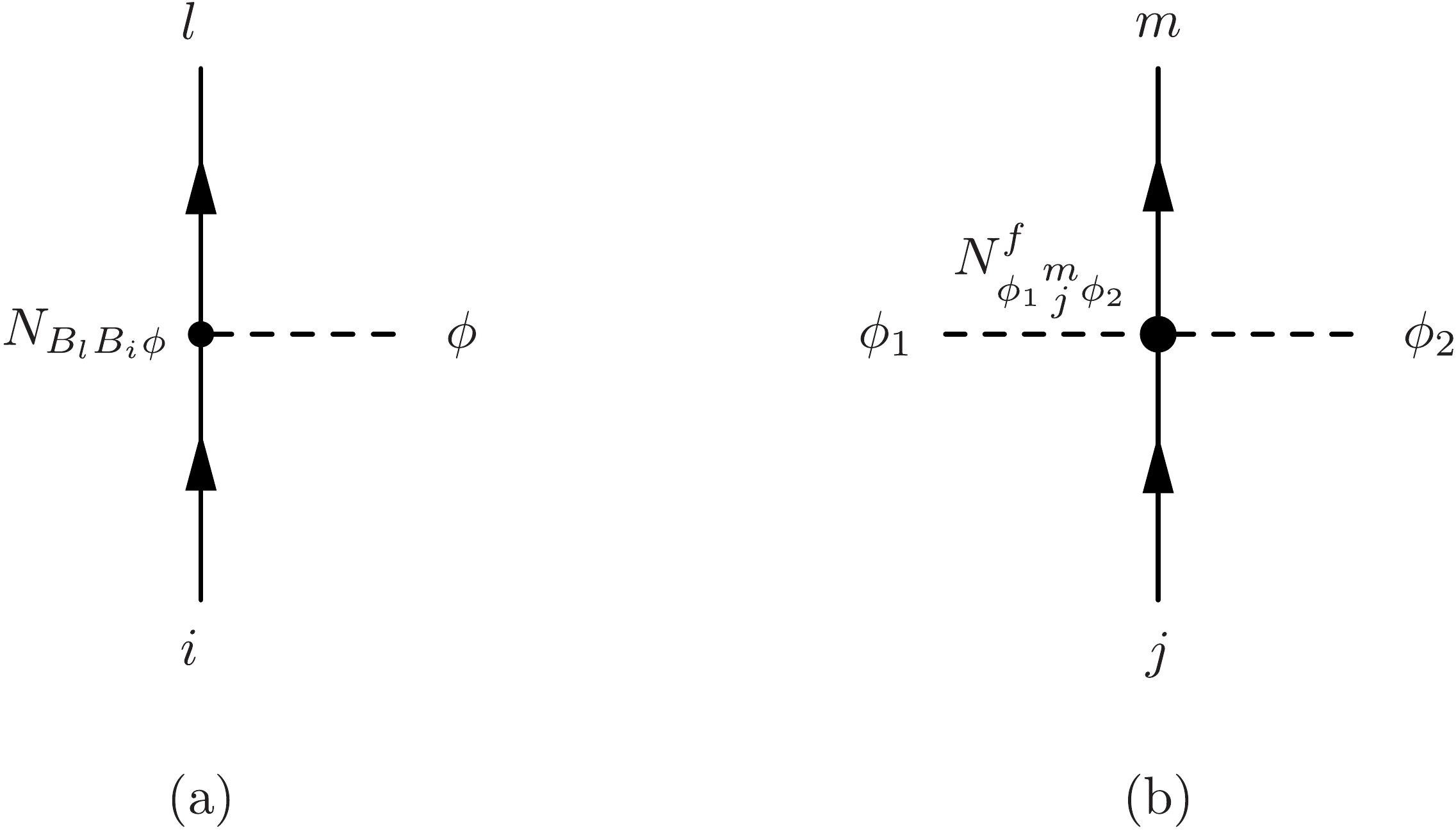}
    \caption{Examples of (a) three-point and (b) four-point vertices
    involving the coefficients $N_{B_lB_i\phi}$ and $N^f_{\phi_1\substack{m\\j}\phi_2}$, respectively.
    Baryons and mesons are denoted by the solid and dashed lines, respectively.}
    \label{fig:2meson}
\end{figure}

To define the isospin-basis potential, we adopt the convention that, in the initial (final) state, index 1 (4) labels the $\Xi$, and indices 2 and 3 (5 and 6) label the nucleons.
This convention is also applied to other terms.
Then, by substituting the thus-obtained potential for $v_{ijk,lmn}$ in Eq.~\eqref{Visospin},
the isospin-basis potential of the TPE term is derived as
\begin{align}
    V_{\mathrm{TPE}}^{\Xi NN}
    &= \mathcal{A}_{23}\left(Y_{123}^{456}+Y_{231}^{564}+Y_{312}^{645}\right),
    \label{eq:TPE}
\end{align}
with
\begin{align}
    Y_{123}^{456}
    &=
    \dfrac{g_Ag_B}{12f_\pi^4}\dfrac{\left(\bm{\sigma}_1\cdot\bm{q}_{41}\right) \left(\bm{\sigma}_3\cdot\bm{q}_{63}\right)}{\left(q_{41}^2+m_\pi^2\right)\left(q_{63}^2+m_\pi^2\right)}
    \left[\bm{\tau}_1\cdot\bm{\tau}_3\left(4c_1 m_\pi^2-2c_3\bm{q}_{41}\cdot\bm{q}_{63}\right)
    +c_4\left\{\bm{\tau}_3\cdot(\bm{\tau}_1\times\bm{\tau}_2)\right\}\left\{\bm{\sigma}_2\cdot(\bm{q}_{41}\times\bm{q}_{63})\right\}\right],
    \label{Y123456}\\
    Y_{231}^{564}
    &=
    \dfrac{g_Ag_B}{12f_\pi^4}\dfrac{\left(\bm{\sigma}_2\cdot\bm{q}_{52}\right)\left(\bm{\sigma}_1\cdot\bm{q}_{41}\right)}{\left(q_{52}^2+m_\pi^2\right)\left(q_{41}^2+m_\pi^2\right)}
    \left[\bm{\tau}_2\cdot\bm{\tau}_1\left(4c_1 m_\pi^2-2c_3\bm{q}_{52}\cdot\bm{q}_{41}\right)
    +c_4\left\{\bm{\tau}_3\cdot(\bm{\tau}_1\times\bm{\tau}_2)\right\}\left\{\bm{\sigma}_3\cdot(\bm{q}_{52}\times\bm{q}_{41})\right\}\right],
    \label{Y231564}\\
    Y_{312}^{645}
    &=
    \dfrac{g_A^2}{12f_\pi^4}\dfrac{\left(\bm{\sigma}_3\cdot\bm{q}_{63}\right)\left(\bm{\sigma}_2\cdot\bm{q}_{52}\right)}{\left(q_{63}^2+m_\pi^2\right)\left(q_{52}^2+m_\pi^2\right)}
    \left[\bm{\tau}_3\cdot\bm{\tau}_2\left(-12u_1 m_\pi^2+6u_3\bm{q}_{63}\cdot\bm{q}_{52}\right)
    -u_4\left\{\bm{\tau}_3\cdot(\bm{\tau}_1\times\bm{\tau}_2)\right\}\left\{\bm{\sigma}_1\cdot(\bm{q}_{63}\times\bm{q}_{52})\right\}\right].
    \label{Y312645}
\end{align}
Here, $\mathcal{A}_{ij}$ is the two-body antisymmetrizer defined by
\begin{align}
    \mathcal{A}_{ij}
    =
    \frac{1}{2}\left(1-\mathcal{P}_{ij}\right),
    \label{antisymmetrizer}
\end{align}
with the exchange operator ${\mathcal{P}_{ij} = \mathcal{P}^{(q)}_{ij}\mathcal{P}^{(\sigma)}_{ij}\mathcal{P}^{(\tau)}_{ij}}$, 
where $\mathcal{P}^{(q)}_{ij}$, $\mathcal{P}^{(\sigma)}_{ij}$, and $\mathcal{P}^{(\tau)}_{ij}$, respectively exchange the parity, spin, and isospin of the subsystem formed by baryons $i$ and $j$.
The pion-decay constant with the finite pion mass is expressed by $f_\pi$, for which we employ $f_\pi = 92.4$~MeV, and $m_\pi=138.04$~MeV is the average pion mass.
The axial-vector coupling constants $g_A$ and $g_B$ are expressed as {$g_A=D+F$} and {$g_B=-D+F$} with {$D \approx 0.8$} and {$F \approx 0.5$}~\cite{PhysRevC.93.014001}.
The LECs $c_i$ and $u_i$ are given as a linear combination of other LECs;
\begin{equation}
\begin{gathered}
    c_1=\dfrac{1}{2}\left(2b_0+b_D+b_F\right),\quad c_3=b_1+b_2+b_3+2b_4,\quad c_4=4\left(d_1+d_2\right),\\
    u_1=\dfrac{1}{2}\left(2b_0+b_D-b_F\right),\quad u_3=b_1+b_2-b_3+2b_4,\quad u_4=4\left(d_1-d_2\right).
\end{gathered}
\label{LECsTPE}
\end{equation}

\subsubsection{One-pion exchange plus contact term}
\label{sec:form_XiNNpot_OPE}
Within the particle-basis representation, the one-meson exchange plus contact (OME) potential for general three-baryon systems can be written as~\cite{PhysRevC.93.014001}
\begin{align}
    v_{\mathrm{OME}}
    &=
    \dfrac{1}{2f_0^2}\dfrac{\bm{\sigma}_i\cdot\bm{q}_{li}}{q_{li}^2+m_{\phi}^2}\left[N_1\bm{\sigma}_k\cdot\bm{q}_{li}+N_2i\left(\bm{\sigma}_j\times\bm{\sigma}_k\right)\cdot\bm{q}_{li}\right],
    \label{v1meson}\\
    N_1&=N_{B_lB_i\phi}\sum_{f=1}^{10}D_fN^f_{\substack{m\\j}\substack{n\\k}\bar{\phi}},
    \label{N1_1meson}\\
    N_2&=N_{B_lB_i\phi}\sum_{f=11}^{14}D_fN^f_{\substack{m\\j}\substack{n\\k}\bar{\phi}},
    \label{N2_1meson}
\end{align}
where $D_f$ are the LECs associated with the four-baryon vertex with one meson as shown by Fig.~\ref{fig:1meson}.

The SU(3) coefficient $N^f_{\substack{m\\j}\substack{n\\k}\phi}$ can be calculated by comparing the minimal nonrelativistic chiral Lagrangian for the diagram illustrated in Fig.~\ref{fig:1meson} in matrix notation and its equivalent expression in the particle basis~\cite{PhysRevC.93.014001}.
The resulting values are given in Appendix~\ref{sec:vertexcoef}.
Note that $N_{B_lB_i\phi}$ is identical to that in the TME potential.

\begin{figure}[!t]
    \centering
    \includegraphics[width=0.2\linewidth]{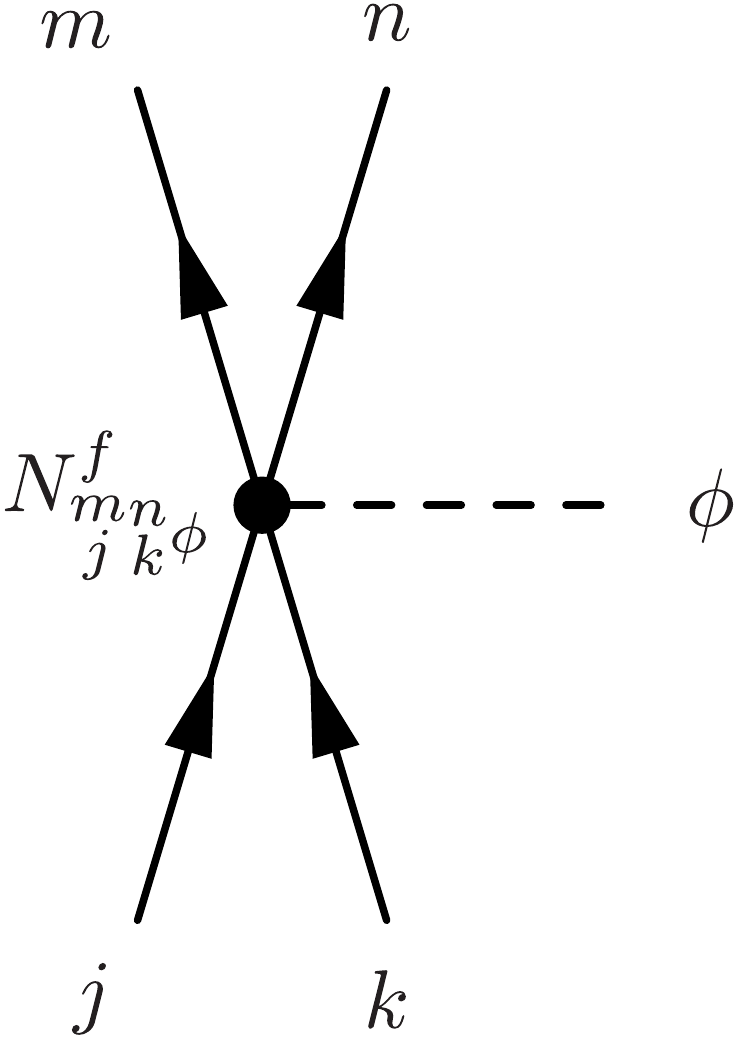}
    \caption{Diagram of a four-baryon vertex including one meson,
    labeled by the coefficient $N^f_{\substack{m\\j}\substack{n\\k}\phi}$.
    The notation is similar to that in Fig.~\ref{fig:2meson}.}
    \label{fig:1meson}
\end{figure}

Then, using Eq.~\eqref{Visospin}, we derive the $\Xi NN$ potential of the OPE term in the isospin-basis representation as
\begin{align}
    V_{\mathrm{OPE}}^{\Xi NN}
    &= X_{123}^{456}+X_{231}^{564}+X_{312}^{645}+\mathcal{P}_{23}\mathcal{P}_{12}X_{312}^{564}+\mathcal{P}_{23}\mathcal{P}_{13}X_{231}^{645},
    \label{eq:OPE}
\end{align}
with
\begin{align}
    X_{123}^{456}
    &=
    \dfrac{g_Bd}{2f_\pi^2}\dfrac{\bm{\sigma}_1\cdot\bm{q}_{41}}{q_{41}^2+m_\pi^2}
    \left[(\bm{\tau}_2-\bm{\tau}_3)\cdot\bm{\tau}_1\,(\bm{\sigma}_2-\bm{\sigma}_3)\cdot\bm{q}_{41}+(\bm{\tau}_1\times\bm{\tau}_2)\cdot\bm{\tau}_3\,(\bm{\sigma}_2\times\bm{\sigma}_3)\cdot\bm{q}_{41}\right],
    \label{X123456}\\
    X_{231}^{564}
    &=
    \dfrac{g_A}{2f_\pi^2}\dfrac{\bm{\sigma}_2\cdot\bm{q}_{52}}{q_{52}^2+m_\pi^2}
    \left[(e_1\bm{\sigma}_3+e_2\bm{\sigma}_1)\cdot\bm{q}_{52}\,\bm{\tau}_2\cdot\bm{\tau}_3
    +(e_3\bm{\sigma}_3+e_4\bm{\sigma}_1)\cdot\bm{q}_{52}\,\bm{\tau}_1\cdot\bm{\tau}_2-e_5(\bm{\tau}_2\times\bm{\tau}_3)\cdot\bm{\tau}_1\,(\bm{\sigma}_3\times\bm{\sigma}_1)\cdot\bm{q}_{52}\right],
    \label{X231564}\\
    X_{312}^{645}
    &=
    \dfrac{g_A}{2f_\pi^2}\dfrac{\bm{\sigma}_3\cdot\bm{q}_{63}}{q_{63}^2+m_\pi^2}\left[(e_4\bm{\sigma}_1+e_3\bm{\sigma}_2)\cdot\bm{q}_{63}\,\bm{\tau}_3\cdot\bm{\tau}_1+(e_2\bm{\sigma}_1+e_1\bm{\sigma}_2)\cdot\bm{q}_{63}\,\bm{\tau}_2\cdot\bm{\tau}_3-e_5(\bm{\tau}_3\times\bm{\tau}_1)\cdot\bm{\tau}_2\,(\bm{\sigma}_1\times\bm{\sigma}_2)\cdot\bm{q}_{63}\right],
    \label{X312645}\\
    X_{312}^{564}
    &=
    \dfrac{g_A}{2f_\pi^2}\dfrac{\bm{\sigma}_3\cdot\bm{q}_{63}}{q_{63}^2+m_\pi^2}
    \left[\{f_1\bm{\sigma}_1+f_2\bm{\sigma}_2+f_3i(\bm{\sigma}_1\times\bm{\sigma}_2)\}\cdot\bm{q}_{63}\,\bm{\tau}_3\cdot\bm{\tau}_1+\{f_2\bm{\sigma}_1+f_1\bm{\sigma}_2+f_3i(\bm{\sigma}_1\times\bm{\sigma}_2)\}\cdot\bm{q}_{63}\,\bm{\tau}_2\cdot\bm{\tau}_3\right.
    \notag\\
    &\left.+\{f_4(\bm{\sigma}_1+\bm{\sigma}_2)+if_5(\bm{\sigma}_1\times\bm{\sigma}_2)\}\cdot\bm{q}_{63}\,i(\bm{\tau}_3\times\bm{\tau}_1)\cdot\bm{\tau}_2\right],
    \label{X312564}\\
    X_{231}^{645}
    &=
    \dfrac{g_A}{2f_\pi^2}\dfrac{\bm{\sigma}_2\cdot\bm{q}_{52}}{q_{52}^2+m_\pi^2}
    \left[\{f_1\bm{\sigma}_3+f_2\bm{\sigma}_1-f_3i(\bm{\sigma}_3\times\bm{\sigma}_1)\}\cdot\bm{q}_{52}\,\bm{\tau}_2\cdot\bm{\tau}_3+\{f_2\bm{\sigma}_3+f_1\bm{\sigma}_1-f_3i(\bm{\sigma}_3\times\bm{\sigma}_1)\}\cdot\bm{q}_{52}\,\bm{\tau}_1\cdot\bm{\tau}_2\right.\notag\\
    &\left.-\{f_4(\bm{\sigma}_3+\bm{\sigma}_1)-if_5(\bm{\sigma}_3\times\bm{\sigma}_1)\}\cdot\bm{q}_{52}\,i(\bm{\tau}_2\times\bm{\tau}_3)\cdot\bm{\tau}_1\right].
    \label{X231645}
\end{align}
The LECs $d$, $e_i$, and $f_i$ involved above can be written in terms of the other LECs $D_i$;
\begin{align}
    d&=\dfrac{1}{6}\left(D_1-D_3+D_8-D_{10}\right),
    \label{LECd}\\
    e_1&=-\dfrac{1}{2}D_4-D_5-\dfrac{1}{2}D_6+D_{10}+D_{12}-D_{13},
    \label{LECe1}\\
    e_2&=-\dfrac{1}{2}D_4-D_5-\dfrac{1}{2}D_6+D_8-D_{12}+D_{13},
    \label{LECe2}\\
    e_3&=-\dfrac{1}{6}D_4-\dfrac{1}{6}D_6-\dfrac{1}{3}D_7+\dfrac{1}{3}D_9+\dfrac{1}{3}D_{12}-\dfrac{1}{3}D_{14},
    \label{LECe3}\\
    e_4&=-\dfrac{1}{6}D_4-\dfrac{1}{6}D_6-\dfrac{1}{3}D_7-\dfrac{1}{3}D_{12}+\dfrac{1}{3}D_{14},
    \label{LECe4}\\
    e_5&=\dfrac{1}{6}D_4-\dfrac{1}{6}D_6,
    \label{LECe5}\\
    f_1&=\dfrac{1}{6}D_4+\dfrac{1}{2}D_5+\dfrac{1}{2}D_6+\dfrac{1}{6}D_7-\dfrac{1}{4}D_8-\dfrac{1}{12}D_9-\dfrac{1}{4}D_{10},
    \label{LECf1}\\
    f_2&=\dfrac{1}{2}D_4+\dfrac{1}{2}D_5+\dfrac{1}{6}D_6+\dfrac{1}{6}D_7-\dfrac{1}{4}D_8-\dfrac{1}{12}D_9-\dfrac{1}{4}D_{10},
    \label{LECf2}\\
    f_3&=-\dfrac{1}{4}D_8+\dfrac{1}{12}D_9+\dfrac{1}{4}D_{10}+\dfrac{2}{3}D_{12}-\dfrac{1}{2}D_{13}-\dfrac{1}{6}D_{14},
    \label{LECf3}\\
    f_4&=-\dfrac{1}{6}D_4-\dfrac{1}{2}D_5-\dfrac{1}{6}D_6+\dfrac{1}{6}D_7+\dfrac{1}{4}D_8-\dfrac{1}{12}D_9+\dfrac{1}{4}D_{10},
    \label{LECf4}\\
    f_5&=\dfrac{1}{4}D_8+\dfrac{1}{12}D_9-\dfrac{1}{4}D_{10}-\dfrac{1}{3}D_{12}+\dfrac{1}{2}D_{13}-\dfrac{1}{6}D_{14}.
    \label{LECf5}
\end{align}

\subsubsection{Contact term}
\label{sec:form_XiNNpot_ct}
Similar to the other meson-exchange potentials, first, we write the potential for the general three-baryon contact force within the particle-basis representation based on the notation of Ref.~\cite{PhysRevC.93.014001};
\begin{align}
    v_{\mathrm{ct}}
    &=
    V^D+\mathcal{P}_{23}^{(\sigma)}\mathcal{P}_{13}^{(\sigma)}(V^D)_{\substack{4\to5\\5\to6\\6\to4}}+\mathcal{P}_{23}^{(\sigma)}\mathcal{P}_{12}^{(\sigma)}(V^D)_{\substack{4\to6\\5\to4\\6\to5}}-\mathcal{P}_{23}^{(\sigma)}(V^D)_{\substack{4\to4\\5\to6\\6\to5}}-\mathcal{P}_{13}^{(\sigma)}(V^D)_{\substack{4\to6\\5\to5\\6\to4}}-\mathcal{P}_{12}^{(\sigma)}(V^D)_{\substack{4\to5\\5\to4\\6\to6}},
    \label{vct}\\
    V^D
    &=-\sum_{f=1}^{11}\sum_{a=1}^{5}\tilde{t}^{f,a}
    \left(N^{f,a}_{\substack{456\\123}}V_{123}^a+N^{f,a}_{\substack{564\\231}}V_{231}^a+N^{f,a}_{\substack{645\\312}}V_{312}^a+N^{f,a}_{\substack{465\\132}}V_{132}^a+N^{f,a}_{\substack{654\\321}}V_{321}^a+N^{f,a}_{\substack{546\\213}}V_{213}^a\right),
    \label{VD}
\end{align}
where $V_{ijk}^{a}$ is defined by
\begin{align}
    V_{ijk}^{a}
    =
    \begin{dcases}
    1 & (a=1),\\
    \bm{\sigma}_j\cdot\bm{\sigma}_k & (a=2)\\
    \bm{\sigma}_k\cdot\bm{\sigma}_i & (a=3)\\
    \bm{\sigma}_i\cdot\bm{\sigma}_j & (a=4)\\
    i\bm{\sigma}_i\cdot\left(\bm{\sigma}_j\times\bm{\sigma}_k\right) & (a=5).
    \end{dcases}
    \label{vijka}
\end{align}
The coefficient $\tilde{t}^{f,a} N^{f,a}_{\substack{lmn\\ijk}}$ determines the six-baryon contact vertex, as shown in Fig.~\ref{fig:ct}.

By comparing the six-baryon contact Lagrangian in matrix notation and that in particle-basis representation~\cite{PhysRevC.93.014001},
we can obtain the SU(3) coefficient $\tilde{t}^{f,a} N^{f,a}_{\substack{lmn\\ijk}}$,
the explicit values of which are listed in Appendix~\ref{sec:vertexcoef}.
By substituting the thus-obtained potential for $v_{ijk,lmn}$ in Eq.~\eqref{Visospin},
the $\Xi NN$-contact potential is derived as
\begin{align}
    V_{\mathrm{ct}}^{\Xi NN}
    &= E_1+E_2(\bm{\sigma}_2+\bm{\sigma}_3)\cdot\bm{\sigma}_1+E_3\bm{\sigma}_2\cdot\bm{\sigma}_3
    +\left[F_1-F_2(\bm{\sigma}_2-\bm{\sigma}_3)\cdot\bm{\sigma}_1-F_1\bm{\sigma}_2\cdot\bm{\sigma}_3\right]\bm{\tau}_1\cdot\bm{\tau}_2\notag\\
    &+\left[F_1-F_2(\bm{\sigma}_2-\bm{\sigma}_3)\cdot\bm{\sigma}_1-F_1\bm{\sigma}_2\cdot\bm{\sigma}_3\right]\bm{\tau}_1\cdot\bm{\tau}_3
    +\left[G_1-E_2(\bm{\sigma}_2+\bm{\sigma}_3)\cdot\bm{\sigma}_1-\dfrac{1}{3}E_1\bm{\sigma}_2\cdot\bm{\sigma}_3\right]\bm{\tau}_2\cdot\bm{\tau}_3\notag\\
    &+F_2\left(\bm{\sigma}_1\times\bm{\sigma}_2\right)\cdot\bm{\sigma}_3\left(\bm{\tau}_1\times\bm{\tau}_2\right)\cdot\bm{\tau}_3.
    \label{eq:ct}
\end{align}
The LECs $E_i$, $F_i$, and $G_1$ can be expressed as a linear combination of the other LECs $C_i$,
which enter the particle-basis Lagrangian~\cite{PhysRevC.93.014001}:
\begin{align}
    &E_1=\dfrac{3}{8}C_3+\dfrac{9}{8}C_5-3C_7-\dfrac{3}{4}C_8-\dfrac{9}{4}C_{10}
    +\dfrac{3}{2}C_{11}+\dfrac{9}{8}C_{13}+\dfrac{27}{8}C_{14}+3C_{16}+\dfrac{3}{4}C_{17},
    \label{LECE1}\\
    &E_2=-\dfrac{1}{4}C_2+\dfrac{1}{8}C_3+\dfrac{3}{8}C_5+\dfrac{1}{2}C_6-\dfrac{1}{2}C_7+\dfrac{1}{4}C_8+\dfrac{3}{4}C_{10}
    +\dfrac{1}{4}C_{12}-\dfrac{1}{8}C_{13}-\dfrac{3}{8}C_{14}-\dfrac{1}{2}C_{16}+\dfrac{1}{4}C_{17},
    \label{LECE2}\\
    &E_3=\dfrac{1}{2}C_1-\dfrac{1}{2}C_2-\dfrac{1}{8}C_3-\dfrac{3}{8}C_5+2C_7+\dfrac{5}{4}C_8+\dfrac{15}{4}C_{10}
    -C_{11}-\dfrac{3}{2}C_{12}-\dfrac{3}{8}C_{13}-\dfrac{9}{8}C_{14}-2C_{16}-\dfrac{5}{4}C_{17},
    \label{LECE3}\\
    &F_1=-\dfrac{1}{12}C_1+\dfrac{1}{24}C_3+\dfrac{1}{24}C_5-\dfrac{1}{4}C_8-\dfrac{1}{4}C_{10}
    +\dfrac{1}{4}C_{11}+\dfrac{1}{8}C_{13}+\dfrac{1}{8}C_{14}+\dfrac{1}{4}C_{17},
    \label{LECF1}\\
    &F_2=\dfrac{1}{24}C_3+\dfrac{1}{24}C_5-\dfrac{1}{12}C_8-\dfrac{1}{12}C_{10}-\dfrac{1}{24}C_{13}
    -\dfrac{1}{24}C_{14}+\dfrac{1}{6}C_{15}-\dfrac{1}{12}C_{17},
    \label{LECE2}\\
    &G_1=-\dfrac{1}{2}C_1+\dfrac{1}{2}C_2-\dfrac{1}{8}C_3-\dfrac{3}{8}C_5-\dfrac{3}{4}C_8
    -\dfrac{9}{4}C_{10}+\dfrac{3}{2}C_{12}-\dfrac{3}{8}C_{13}-\dfrac{9}{8}C_{14}+\dfrac{3}{4}C_{17}.
    \label{LECG1}
\end{align}
\begin{figure}[!t]
    \centering
    \includegraphics[width=0.2\linewidth]{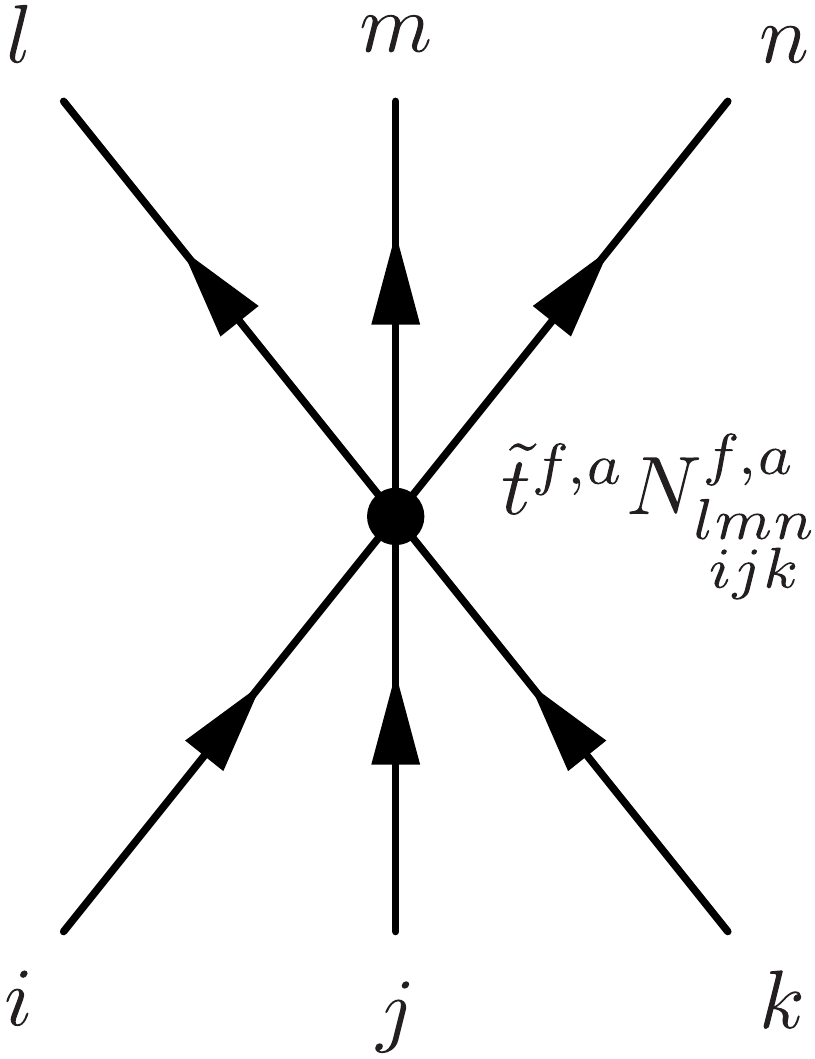}
    \caption{The six-baryon vertex labeled by the coefficient $\tilde{t}^{f,a} N^{f,a}_{\substack{lmn\\ijk}}$.}
    \label{fig:ct}
\end{figure}

\subsection{Deuteron--$\Xi^-$ potential in coordinate space}
\label{sec:form_dXipot}
To incorporate the $\Xi NN$ 3BF formulated in the previous section into calculations of the $d$--$\Xi^-$ correlation function,
we calculate the potential between $d$ and $\Xi^-$ in coordinate space.
In this section, we briefly outline the calculations; 
the details of computing the potential are relegated to Appendix~\ref{sec:dXipot_app}.

The deuteron consists of the proton and neutron,
and hence, we consider the $p$--$n$--$\Xi^-$ three-body system,
for which we introduce the coordinates as illustrated in Fig.~\ref{fig:coordinates}.
The vector $\bm{r}$ ($\bm{R}$) denotes the relative coordinate between $p$ and $n$ (between the center-of-mass of the $p$--$n$ system and $\Xi^-$). 
Using the proton mass $m_2$ and neutron mass $m_3$,
we define the coordinates $\bm{r}_1$, $\bm{r}_2$, and $\bm{r}_3$ in terms of $\bm{r}$ and $\bm{R}$;
\begin{align}
    \bm{r}_1 &= \bm{r},
    \label{r1}\\
    \bm{r}_2 &= \bm{R}-M_2\bm{r},
    \label{r2}\\
    \bm{r}_3 &= \bm{R}+M_3\bm{r},
    \label{r3}
\end{align}
where the mass ratio $M_i$ is defined by $M_i = m_i/(m_2+m_3)$.

The coordinate-space potential $U_{\mathrm{3BF}}^{(\sigma)}$ 
depends on $\sigma$, which is the total spin of the $\Xi NN$ system,
and is given by folding the coordinate-space $\Xi NN$ potential with the deuteron wave function $\varphi$;
\begin{align}
    U_{\mathrm{3BF}}^{(\sigma)}(R)
    &=
    \frac{1}{(4\pi)^2}\iint d\bm{r}d\hat{\bm{R}}
    \frac{\varphi^2(r)}{r^2}
    W^{(\sigma)}(\bm{r},\bm{R}),
    \label{Ufold_general}
\end{align}
with $\hat{\bm{R}}=\bm{R}/R$.
To evaluate $\varphi$, we follow Ref.~\cite{PhysRevC.103.065205},
i.e., the ground state of deuteron is assumed to have the $S$-wave component only
and is computed using the Argonne $V4'$ nucleon--nucleon interaction~\cite{PhysRevLett.89.182501}.
The coordinate-space $\Xi NN$ potential, $W^{(\sigma)}$, is defined by the Fourier transform;
\begin{align}
    W^{(\sigma)}(\bm{r},\bm{R})
    &=
    \frac{1}{(2\pi)^6} \iiint d\bm{q}_1 d\bm{q}_2 d\bm{q}_3
    \exp\left[i(\bm{q}_1\cdot\bm{r}_1 + \bm{q}_2\cdot\bm{r}_2 + \bm{q}_3\cdot\bm{r}_3)\right]
    \delta(\bm{q}_1+\bm{q}_2+\bm{q}_3)
    \notag\\
    &\times
    \Braket{\Upsilon_{\sigma m_\sigma}\Theta_{1/2,-1/2}|
    V_{(0)}^{\Xi NN}(\bm{q}_1,\bm{q}_2,\bm{q}_3)
    |\Upsilon_{\sigma m_\sigma}\Theta_{1/2,-1/2}}.
    \label{WrR}
\end{align}
Here, we simplify the notation of the momentum transfers as
$\bm{q}_1=\bm{q}_{41}$, $\bm{q}_2=\bm{q}_{52}$, and $\bm{q}_3=\bm{q}_{63}$.
Owing to the delta function responsible for the momentum conservation, the integration variables can be chosen arbitrarily from any pair out of the three momentum transfers, depending on the form of $V_{(0)}^{\Xi NN}$, which is the central component of the $\Xi NN$ potential in momentum space.
In the spin--isospin matrix element, $m_\sigma$ is the third component of $\sigma$
and $\Ket{\Upsilon_{\sigma m_\sigma}}$ $\left(\Ket{\Theta_{1/2,-1/2}}\right)$ represents the spin (isospin) state of the $d$--$\Xi^-$ system. 

We take into account only the central component of the $\Xi NN$ 3BF, as it is expected to provide the dominant contribution to the correlation function.
This is justified by the short-range nature of the strong interaction and relatively low-energy scattering relevant to the correlation function (see, e.g., Ref.~\cite{CHO2017279}).

In Appendix~\ref{sec:dXipot_app} we show the explicit form of $V_{(0)}^{\Xi NN}$, the spin--isospin matrix elements involved in Eq.~\eqref{WrR}, and the multipole expansion relevant to Eq.~\eqref{Ufold_general}.

\begin{figure}[!t]
    \centering
    \includegraphics[width=0.4\linewidth]{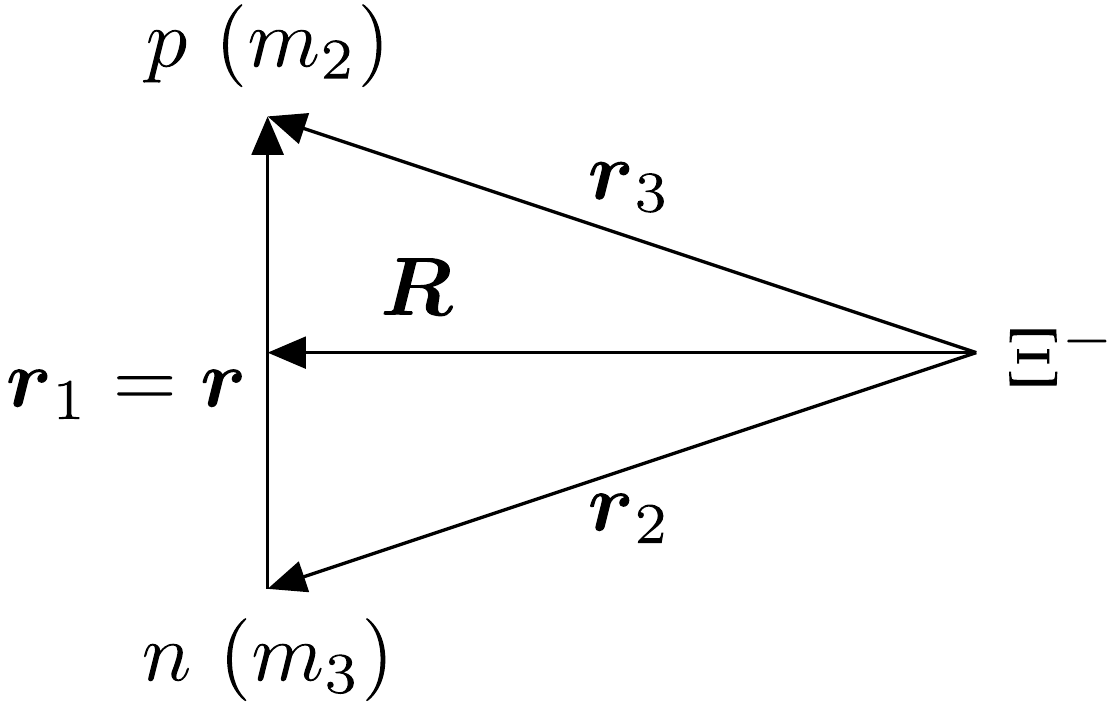}
    \caption{Coordinates of the $p$--$n$--$\Xi^-$ system.
    The proton and neutron masses are represented by $m_2$ and $m_3$, respectively.}
    \label{fig:coordinates}
\end{figure}

As a result of the above procedure, Eq.~\eqref{Ufold_general} is rewritten as 
$U_\mathrm{3BF}^{(\sigma)} = U_\mathrm{TPE}^{(u_1)}+U_\mathrm{TPE}^{(u_3)}+U_\mathrm{OPE}^{(\sigma)}+U_\mathrm{ct}^{(\sigma)}$, with
\begin{align}
    U_\mathrm{TPE}^{(u_1)}(R)
    &=
    \frac{g_A^2 u_1 m_\pi^{6}}{32\pi^2 f_\pi^4}
    \sum_{\lambda_1=0,1} \sum_{\lambda_2=0,1} \sum_{l}
    (-)^{\lambda_1} R^{\lambda_1+\lambda_2}
    \widehat{1-\lambda_1}\widehat{1-\lambda_2}
    \left[\left(\mqty{3 \\ 2\lambda_1}\right) \left(\mqty{3 \\ 2\lambda_2}\right) \right]^{\frac{1}{2}}\notag\\
    &\times(1-\lambda_1,0,1-\lambda_2,0|l0)(\lambda_10\lambda_20|l0)\left\{\mqty{1-\lambda_1 & \lambda_1   & 1\\
           \lambda_2         & 1-\lambda_2 &l }\right\}
    \notag\\
    &\times\int dr \varphi^2(r)(M_3r)^{1-\lambda_1}(M_2r)^{1-\lambda_2}\int_{-1}^{1}dw P_l(w) \left(r_2 r_3\right)^{-1}
    Z_1(m_\pi r_2)f_{R_0}(r_2) Z_1(m_\pi r_3)f_{R_0}(r_3),
    \label{UTPEu1}\\
    U_\mathrm{TPE}^{(u_3)}(R)
    &=-\dfrac{g_A u_3 m_\pi^6}{192\pi^2 f_\pi^4}
    \int dr \varphi^2(r)\int_{-1}^{1}dw Y(m_\pi r_2)f_{R_0}(r_2) Y(m_\pi r_3)f_{R_0}(r_3)
    \notag\\
    &\quad-\dfrac{g_A u_3 m_\pi^6}{96\pi^2 f_\pi^4}
    \sum_{\lambda_1=0}^{2}\sum_{\lambda_2=0}^{2}\sum_{l}(-)^{\lambda_1}R^{\lambda_1+\lambda_2}
    \widehat{2-\lambda_1}\widehat{2-\lambda_2}
    \left[\left(\mqty{5 \\ 2\lambda_1}\right) \left(\mqty{5 \\ 2\lambda_2}\right) \right]^{\frac{1}{2}}
    \notag\\
    &\quad\times
    (2-\lambda_1,0,2-\lambda_2,0|l0)(\lambda_10\lambda_20|l0)
    \left\{\mqty{2-\lambda_1&\lambda_1&2\\ \lambda_2&2-\lambda_2&l}\right\}
    \notag\\
    &\quad\times\int dr \varphi^2(r)(M_3r)^{2-\lambda_1}(M_2r)^{2-\lambda_2}\int_{-1}^{1}dw P_l(w) \left(r_2 r_3\right)^{-2}Z(m_\pi r_2)f_{R_0}(r_2)Z(m_\pi r_3)f_{R_0}(r_3),
    \label{UTPEu3}\\
    U_{\mathrm{OPE}}^{(\sigma)}(R)
    &=
    -\dfrac{g_A G_e^{(\sigma)}m_\pi^3}{16 \pi f_\pi^2}\int dr \varphi^2(r)\int_{-1}^{1}dw\left\{Y(m_\pi r_2)f_{R_0}(r_2)D_{R_0}(r_3)+Y(m_\pi r_3)f_{R_0}(r_3)D_{R_0}(r_2)\right\}
    \notag\\
    &\quad-\dfrac{3 g_A G_f^{(\sigma)}m_\pi^3}{8\pi f_\pi^2}\int dr \varphi^2(r)\int_{-1}^{1}dw\left\{Y(m_\pi r_2)f_{R_0}(r_2)D_{R_0}(r)+Y(m_\pi r_3)f_{R_0}(r_3)D_{R_0}(r)\right\},
    \label{UOPE}\\
    U_{\mathrm{ct}}^{(\sigma)}(R)
    &=
    \dfrac{H^{(\sigma)}}{2}\int dr \varphi^2(r)\int_{-1}^{1}dw D_{R_0}(r_2)D_{R_0}(r_3).
    \label{Uct}
\end{align}
Here, the dependence of $r_2$ and $r_3$ on $w$ is not explicitly shown, but they are expressed as functions of $w$ by
\begin{align}
    r_2=&\sqrt{M_2^2 r^2+R^2-2M_2rR w},
    \label{r2mag}\\
    r_3=&\sqrt{M_3^2 r^2+R^2+2M_3rR w},
    \label{r3mag}
\end{align}
where $w=\cos\theta$ with $\theta$ being the angle between $\bm{r}$ and $\bm{R}$.
The integrations over $r$ and $w$ are evaluated numerically.
Equation~\eqref{UTPEu3} involves the binomial coefficient defined by
\begin{align}
    \left(\mqty{a \\ b}\right)
    =
    \frac{a!}{b!(a-b)!}.
    \label{binomial}
\end{align}
The Wigner $6$-$j$ symbol is expressed by $\left\{\mqty{\cdots\\ \cdots}\right\}$,
and the Legendre polynomial is denoted by $P_l$.
We use the abbreviation $\hat l = \sqrt{2l+1}$.

The functions $Y$, $Z_1$, and $Z$ are defined by
\begin{align}
    Y(x)&=\dfrac{\exp(-x)}{x},
    \label{Yx}\\
    Z_1(x)
    &=
    -\frac{d}{dx}Y(x)
    =
    \left(1 + \frac{1}{x}\right)Y(x),
    \label{Z1x}\\
    Z(x)&=\left(1 +\frac{3}{x} +\frac{3}{x^2}\right)Y(x).
    \label{Zx}
\end{align}
For these functions we employ the following regularization~\cite{PhysRevC.90.054323}.
Namely, the Yukawa functions involved in the long-range terms are regularized as
\begin{align}
    Y(m_\pi r) &\to Y(m_\pi r) f_{R_0}(r),
    \label{regY}\\
    f_{R_0}(r) &= 1-\exp\left[-\left(\dfrac{r}{R_0}\right)^5\right].
    \label{regulator}
\end{align}
Accordingly, the delta functions,
which originate from the contact vertices [see Eqs.~\eqref{WOPE} and~\eqref{Wct} in Appendix~\ref{sec:dXipot_app}], are regularized similarly,
\begin{align}
    \delta(\bm{r}) &\to \delta_{R_0}(\bm{r}) \equiv D_{R_0}(r) = \alpha\exp\left[-\left(\dfrac{r}{R_0}\right)^5\right],
    \label{regdelta}\\
    \alpha &= \dfrac{5}{4\pi\Gamma(3/5)R_0^3},
    \label{deltaalpha}
\end{align}
where $\Gamma$ is the Gamma function and the normalization constant $\alpha$ ensures that 
\begin{align}
    \int d \bm{r}\delta_{R_0}(\bm{r})=1.
    \label{deltanorm}
\end{align}
In Eqs.~\eqref{regulator} and~\eqref{regdelta}, the exponent 5 was adopted as the smallest value that makes the integral over $r$ convergent for the calculation of $U_\mathrm{3BF}^{(\sigma)}$.
We employ $R_0 = 1.0$~fm~\cite{PhysRevC.90.054323},
while also varying it by 10\% to examine its effect on the correlation functions.

The newly defined LECs $G_e^{(\sigma)}$, $G_f^{(\sigma)}$ and $H^{(\sigma)}$ in Eqs.~\eqref{UOPE} and~\eqref{Uct} are expressed in terms of the LECs given in Sec.~\ref{sec:form_XiNNpot}:
\begin{align}
    G_e^{(\sigma)}
    &=
    \begin{dcases}
        -e_1+2e_2 &(\sigma=1/2)\\
        -e_1-e_2 &(\sigma=3/2),
    \end{dcases}
    \label{G1sigma}\\
    G_f^{(\sigma)}
    &=
    \begin{dcases}
        \dfrac{-f_1-f_2-6f_3+2f_4+6f_5}{8} &(\sigma=1/2)\\
        \dfrac{f_1+f_2-2f_4}{4} &(\sigma=3/2),
    \end{dcases}
    \label{G2sigma}\\
    H^{(\sigma)}
    &=
    \begin{dcases}
        2E_1-16E_2+E_3-3G_1 &(\sigma=1/2)\\
        2E_1+8E_2+E_3-3G_1 &(\sigma=3/2).
    \end{dcases}
    \label{H1sigma}
\end{align}

\end{widetext}

\subsection{Decuplet saturation approximation}
\label{sec:form_DSA}
To determine the LECs involved in $U_{\mathrm{3BF}}^{(\sigma)}$, we employ the DSA~\cite{PETSCHAUER2017347}, 
which assumes that the dominant contribution arises from intermediate excitations of octet baryons to decuplet baryons during the three-baryon interaction.
By incorporating this mechanism into the contact Lagrangian $\mathcal{L}_{BBBB^*}$ that involves three octet baryons and one decuplet baryon~\cite{PETSCHAUER2017347,PhysRevC.74.064002}, one can approximate the LECs.

As a result of the DSA, the LECs $u_1$ and $u_3$ of the TPE term are approximated as
\begin{align}
    u_1 &\approx 0,
    \label{u1DSA}\\
    u_3
    &\approx
    -\dfrac{2C^2}{9\Delta}.
    \label{u3DSA}
\end{align}
Here $\Delta$ denotes the average decuplet--octet baryon mass splitting, 
and we take $\Delta = 203$~MeV, which is evaluated using the masses of the baryons considered, namely, the proton, neutron, $\Delta(1232)$, $\Xi^0$, $\Xi^-$, and $\Xi(1530)$~\cite{PhysRevD.110.030001}.
We employ the large-$N_c$ value for the LEC $C$,
namely, ${C=3g_A/4}$~\cite{PETSCHAUER2017347,KAISER1998395,VonkEPJA2025}.
Then, under the DSA, the LECs $G_e^{(\sigma)}$ and $G_f^{(\sigma)}$, 
which enter $U_{\mathrm{OPE}}^{(\sigma)}$, become independent of $\sigma$:
\begin{align}
    G_e^{(\sigma)}
    &\approx
    \dfrac{CG^\prime}{9f_\pi^2\Delta},
    \label{GeDSA}\\
    G_f^{(\sigma)}
    &\approx
    \dfrac{CG^\prime}{36f_\pi^2\Delta},
    \label{GfDSA}\\
    G^\prime
    &=
    f_\pi^2\left(2H_1+4H_2\right).
    \label{Gprime}
\end{align}
Similarly, the DSA simplifies the contact LEC $H^{(\sigma)}$ as
\begin{align}
    H^{(\sigma)}
    &\approx
    \dfrac{H^\prime}{9f_\pi^4\Delta},
    \label{HDSA}\\
    H^\prime
    &=
    f_\pi^4\left(4H_1^2+24H_1H_2+24H_2^2\right).
    \label{Hprime}
\end{align}
Since $H_1$ and $H_2$, which are the LECs involved in $\mathcal{L}_{BBBB^*}$~\cite{PETSCHAUER2017347}, enter the potential only through $G^\prime$ and $H^\prime$,
it is sufficient to determine $G^\prime$ and $H^\prime$ without fixing $H_1$ and $H_2$ individually.

An important consequence of the DSA is that the number of LECs in 
$U_{\mathrm{OPE}}^{(\sigma)}$ and $U_{\mathrm{ct}}^{(\sigma)}$ is greatly reduced to a single LEC in each, $G^\prime$ and $H^\prime$, respectively. 
In the following calculations, we vary these dimensionless LECs within the range 
${-1 \leq G^\prime \leq 1}$ and ${-1 \leq H^\prime \leq 1}$, 
so as not to significantly change their orders of magnitude.
Note that the range of variation of $G^\prime$ and $H^\prime$ provides only a rough estimate of the $\Xi NN$ 3BF.
The same range was also adopted in Ref.~\cite{PETSCHAUER2017347}, where a contact LEC for the $\Lambda NN$ 3BF was varied accordingly.

Equation~\eqref{u3DSA} provides another important insight that, under the DSA, the $\Xi NN$-TPE contribution is considerably smaller than TPE three-nucleon forces.
This is because a TPE LEC of three-nucleon forces corresponding to $u_3$ is given by ${c_3 \approx -{8C^2/(9\Delta)}}$~\cite{PETSCHAUER2017347}, 
which is four times greater than Eq.~\eqref{u3DSA}.
This $c_3$-dependent three-nucleon force with the DSA is known to be the so-called Fujita--Miyazawa three-nucleon force~\cite{10.1143/PTP.17.360}.

\subsection{Correlation function}
\label{sec:form_CF}
To compute the $d$--$\Xi^-$ correlation function $C_{d\Xi^-}$, we employ an approach similar to that of Ref.~\cite{PhysRevC.103.065205}, with two main differences;
(i) we take into account the $\Xi NN$ 3BF, and (ii) we disregard the effect of deuteron breakup, which was found to give only minor contributions.
Thus, following Refs.~\cite{KOONIN197743,PhysRevD.33.1314}, $C_{d\Xi^-}$ is given by
\begin{align}
    C_{d\Xi^-}(q)
    &=
    4\pi \int dR R^2 \mathcal{S}(R)
    \sum_{L=1} \hat L^2 
    \left[ \frac{F_L(q R)}{q R}\right]^2
    \notag\\
    &+\frac{2\pi}{3}
    \int dR R^2 \mathcal{S}(R)
    \sum_{\sigma=\frac{1}{2},\frac{3}{2}} \hat \sigma^2 
    \left| \frac{\chi_{L=0}^{(\sigma)}(q,R)}{q R}\right|^2,
    \label{eq:CF}
\end{align}
where $F_L$ is the regular Coulomb wave function with the orbital angular momentum $L$
and $q$ is the relative momentum between $d$ and $\Xi^-$.
The source function $\mathcal{S}$ used here is the same as that employed in Ref.~\cite{PhysRevC.103.065205}:
\begin{align}
    \mathcal{S}(R)=\dfrac{1}{(4\pi b^2)^{3/2}}\exp\left(-\dfrac{R^2}{4b^2}\right),
    \label{sourcefunc}
\end{align}
with the source size $b$. 

The $S$-wave component of the radial distorted wave $\chi_{L=0}^{(\sigma)}$ satisfies the Schr\"odinger equation,
\begin{align}
    \left[-\frac{1}{2\mu_R}\frac{d^2}{dR^2}+U^{(\sigma)}(R)-E\right]
    \chi_{L=0}^{(\sigma)}(q,R)
    =0,
    \label{Schreq}
\end{align}
with the boundary condition,
\begin{align}
    \chi_{L=0}^{(\sigma)}(q,R)
    \to
    \frac{i}{2}\left[\mathcal{U}_{L=0}^{(-)}(qR) - S^{(\sigma)}\mathcal{U}_{L=0}^{(+)}(qR) \right].
    \label{bc}
\end{align}
Here, $\mu_R$ is the reduced mass of the $d$--$\Xi^-$ system, 
and $S^{(\sigma)}$ is the scattering matrix.
The $S$-wave component of the Coulomb wave functions $\mathcal{U}_{L=0}^{(-)}$ and $\mathcal{U}_{L=0}^{(+)}$ have the incoming and outgoing boundary conditions, respectively.
In Eq.~\eqref{Schreq}, using $U_\mathrm{3BF}^{(\sigma)}$ formulated in Sec.~\ref{sec:form_dXipot}, the distorting potential $U^{(\sigma)}$ is decomposed into
\begin{align}
    U^{(\sigma)}(R) = U_\mathrm{2BF}^{(\sigma)}(R)+U_\mathrm{3BF}^{(\sigma)}(R)+V_\mathrm{C}(R).
    \label{distpot}
\end{align}
Note that, owing to Eqs.~\eqref{GeDSA},~\eqref{GfDSA}, and~\eqref{HDSA}, $U_\mathrm{3BF}^{(\sigma)}$ no longer depends on $\sigma$ in actual calculations under the DSA.
The distorting potential $U_\mathrm{2BF}^{(\sigma)}$ originates from $\Xi N$ two-baryon forces (2BFs) and $V_\mathrm{C}$ is the Coulomb potential with respect to the $d$--$\Xi^-$ system.
These potentials are taken from Ref.~\cite{PhysRevC.103.065205},
i.e., $U_\mathrm{2BF}^{(\sigma)}$ is obtained by folding the $\Xi N$ 2BF derived through the HAL QCD method~\cite{SASAKI2020121737}
with the deuteron wave function, which is consistently used for $U_\mathrm{3BF}^{(\sigma)}$,
and $V_\mathrm{C}$ is expressed as
\begin{align}
    V_\mathrm{C}(R)
    =
    \begin{dcases}
        -\frac{e^2}{2R_{\mathrm{C}}}\left(3-\frac{R^2}{R_{\mathrm{C}}^2}\right) & (R\leq R_{\mathrm{C}})\\
        -\frac{e^2}{R_{\mathrm{C}}} & (R > R_{\mathrm{C}}),
    \end{dcases}
\end{align}
with $R_{\mathrm{C}}=1.5$~fm.

\section{Results}
\label{sec:res}
\subsection{Comparison of potentials}
\label{sec:respot}
\begin{figure}[!t]
    \includegraphics[width=1.0\linewidth]{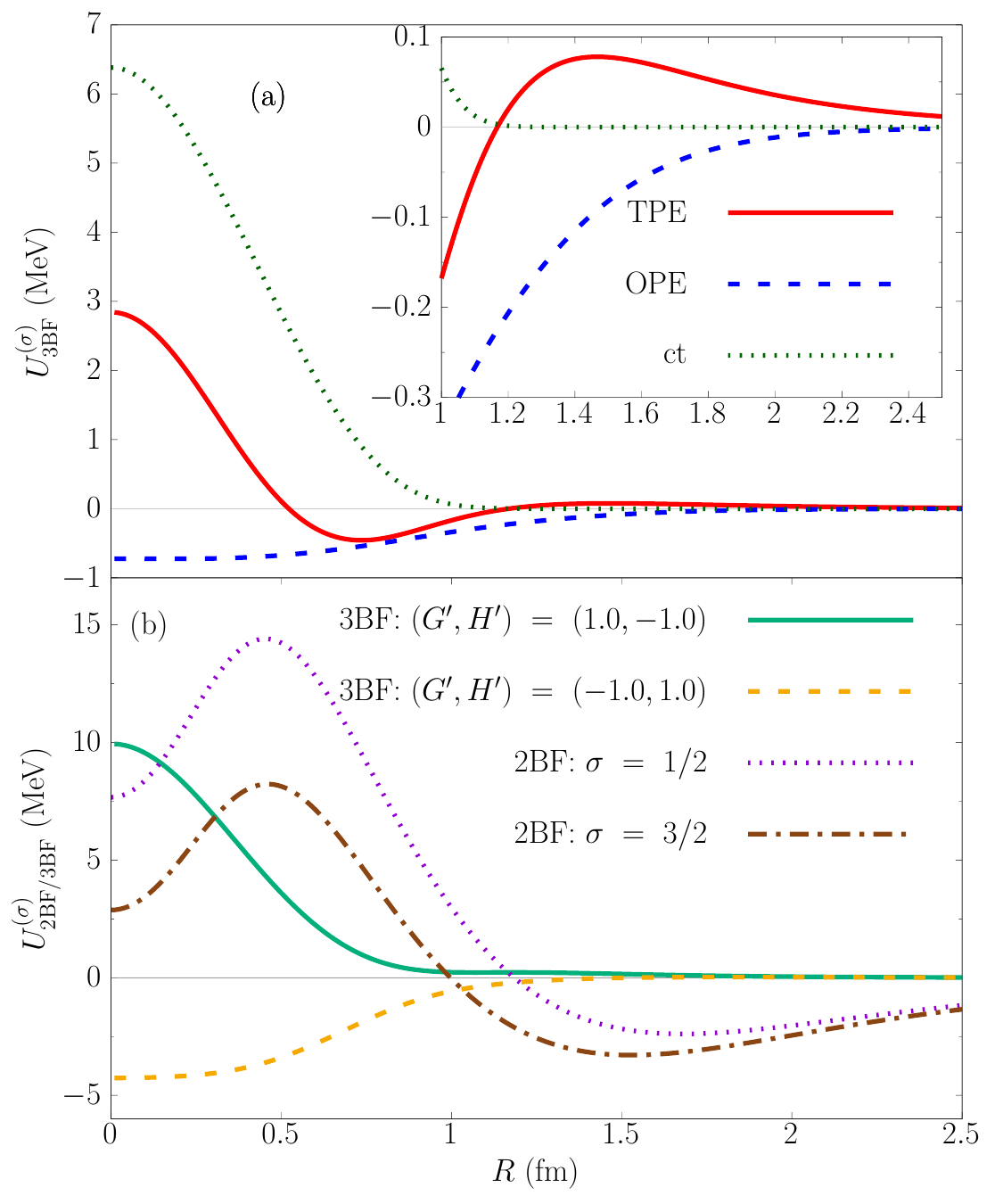}
    \caption{(a) The potentials $U_{\mathrm{TPE}}^{(u_3)}$ (solid line), $U_{\mathrm{OPE}}^{(\sigma)}$ (dashed line), and $U_{\mathrm{ct}}^{(u_3)}$ (dotted line) as a function of $R$.
    (b) Comparison of $U_{\mathrm{3BF}}^{(\sigma)}$ with $U_{\mathrm{2BF}}^{(\sigma)}$.
    The two sets of the LECs $G'$ and $H'$ are employed. See the text for more detail.
    }
    \label{fig:pot}
\end{figure}

Figure~\ref{fig:pot}(a) shows $U_{\mathrm{TPE}}^{(u_3)}$, $U_{\mathrm{OPE}}^{(\sigma)}$, and $U_{\mathrm{ct}}^{(\sigma)}$ as a function of $R$,
and is depicted by the solid, dashed, and dotted lines, respectively.
Note that $U_{\mathrm{TPE}}^{(u_1)}$ defined by Eq.~\eqref{UTPEu1} becomes zero under the DSA.
As a representative case, we fix both ${G^\prime=1}$ and ${H^\prime=1}$, with the cutoff ${R_0=1.0}$~fm. One sees that the contact term can have the largest magnitude at ${R \lesssim 1.0}$~fm. 
Beyond this range, the TPE and OPE terms become dominant
although their absolute magnitude is much smaller compared with that of the contact term near the origin. 

In Fig.~\eqref{fig:pot}(b), we compare $U_{\mathrm{3BF}}^{(\sigma)}$ with $U_{\mathrm{2BF}}^{(\sigma)}$.
In this case, we select two sets of the LECs, one giving the most attractive and the other the most repulsive potential, in order to maximize the variation in $U_{\mathrm{3BF}}^{(\sigma)}$.
These parameter sets correspond to the solid and dashed lines, with ${(G',H')=(1.0,-1.0)}$ and ${(G',H')=(-1.0,1.0)}$, respectively.
The dotted and dash-dotted lines are $U_{\mathrm{2BF}}^{(\sigma)}$ with $\sigma=1/2$ and $3/2$, respectively.
In both sets of the LECs, $U_{\mathrm{3BF}}^{(\sigma)}$ are found to be weaker in magnitude and shorter in range compared with $U_{\mathrm{2BF}}^{(\sigma)}$.

\subsection{Effect of $\Xi NN$ three-baryon force on correlation function}
\label{sec:resCF}
\begin{figure}[!t]
    \centering
    \includegraphics[width=1.0\linewidth]{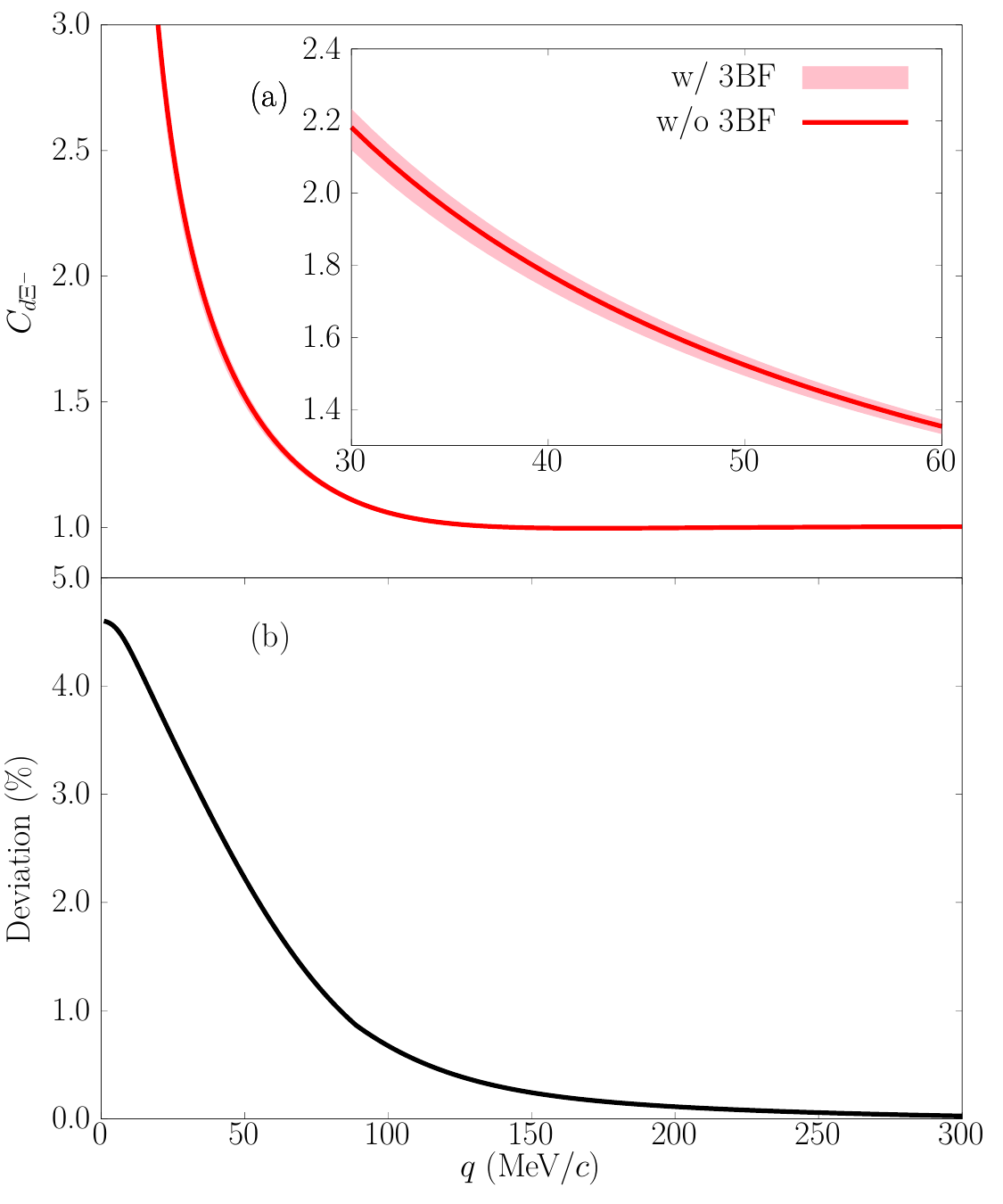}
    \caption{(a) The $d$--$\Xi^-$ correlation function. The horizontal axis is the relative momentum $q$ between $d$ and $\Xi^-$. The solid line is obtained by neglecting the 3BF.
    A band is plotted with its edges corresponding to the most attractive and most repulsive $U_{\mathrm{3BF}}^{(\sigma)}$. 
    (b) The effect of the 3BF on the correlation function, quantified as the relative deviation of the band boundaries from the results neglecting the 3BF.
    }
    \label{fig:CF}
\end{figure}
Figure~\ref{fig:CF}(a) represents the calculated result of the correlation function $C_{d\Xi^-}$ as a function of $q$. 
We vary the LECs within the range $-1 \leq G^\prime \leq 1$ and $-1 \leq H^\prime \leq 1$, 
and thus, the most attractive and most repulsive cases are taken as the boundaries, forming a band.
The solid line corresponds to the correlation function calculated without the $\Xi NN$ 3BF.
The cutoff and source size are set to be ${R_0=1.0}$~fm and ${b=1.2}$~fm, respectively. 
To quantify the effect of the 3BF, we present in Fig.~\ref{fig:CF}(b) the maximum relative deviation of the band boundaries from the central (solid) line at each $q$. 
As seen in these two figures, the effect of the 3BF on $C_{d\Xi^-}$ are very limited, at most around $4\%$. 
This is even smaller than the effect of deuteron breakup, which was found to be about $6\%$--$8\%$~\cite{PhysRevC.103.065205}. 
One possible reason for the small 3BF effect on the correlation function is the short-range nature of the 3BFs.
As shown in Fig.~\ref{fig:pot}, $U_{\mathrm{3BF}}^{(\sigma)}$ becomes nearly zero around $R\gtrsim 1.0$~fm, in contrast to $U_{\mathrm{2BF}}^{(\sigma)}$.
We numerically confirmed that $C_{d\Xi^-}$ is mainly sensitive to the potential in the range $1.0 \lesssim R \lesssim 2.0$~fm, although it depends on $b$.

\subsection{Variation of parameters}
\label{sec:respara}
\begin{figure}[!t]
    \centering
    \includegraphics[width=1.0\linewidth]{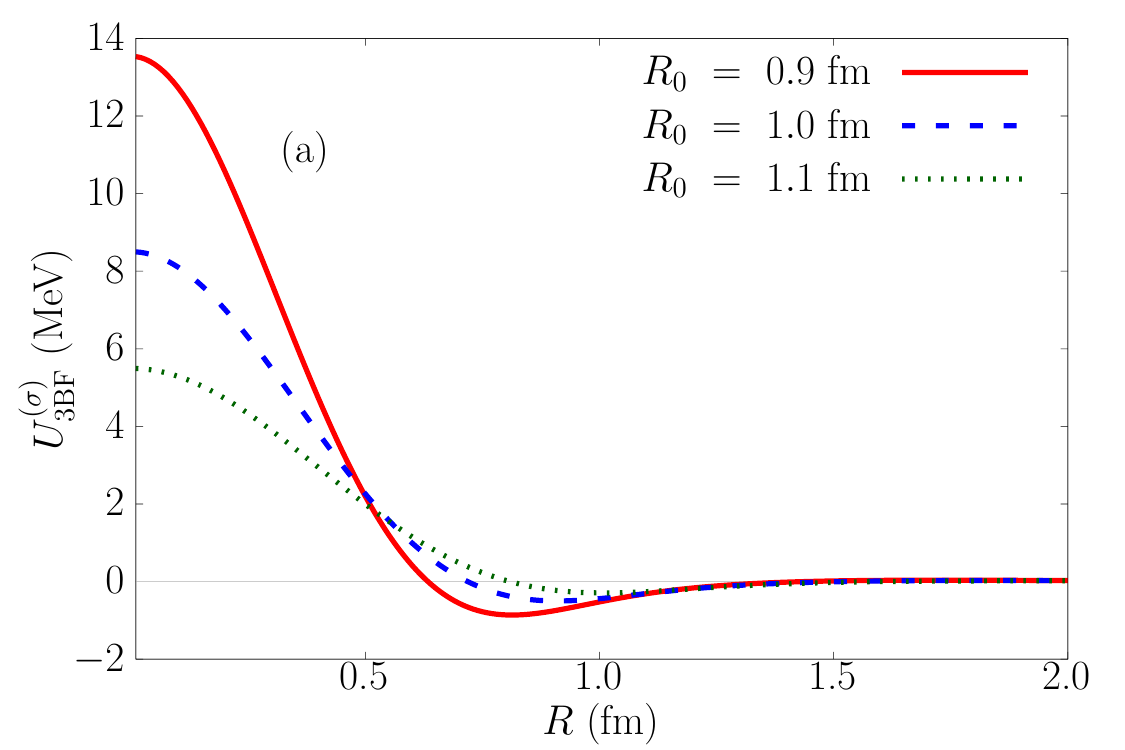}
    \includegraphics[width=1.0\linewidth]{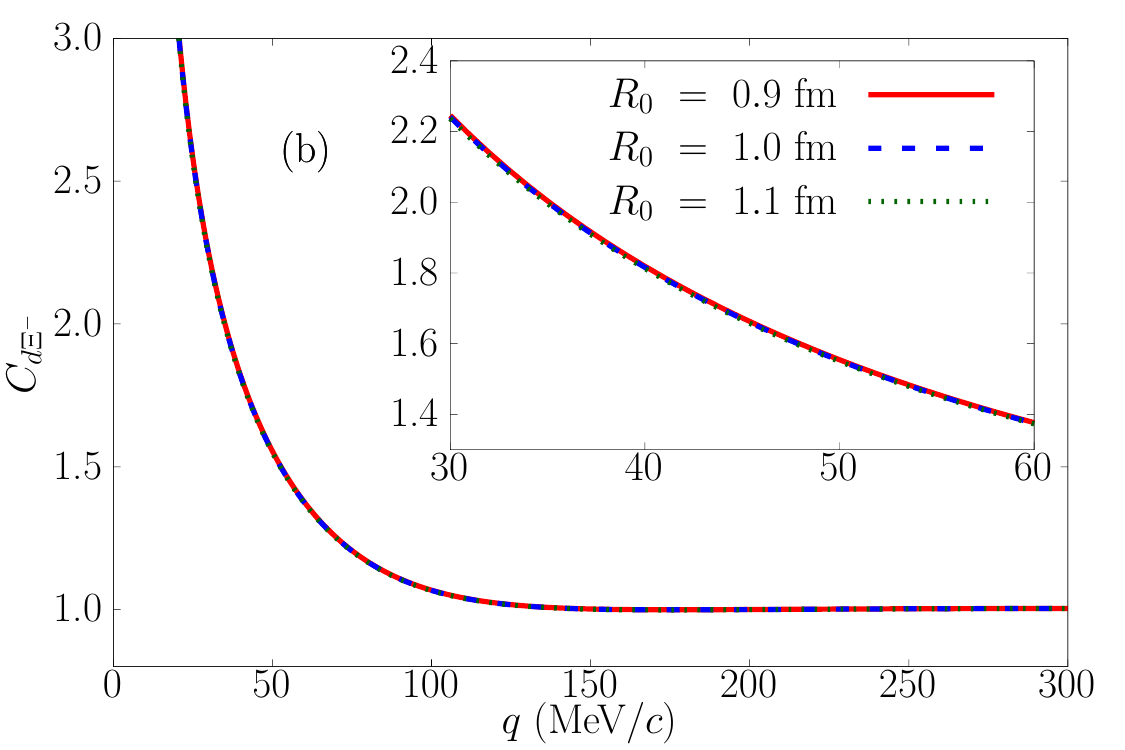}
    \caption{(a) Potentials and (b) the correlation function for different values of the cutoff $R_0$. 
    We fix the LECs and source size as ${G'=1}$, ${H'=1}$, and ${b=1.2}$~fm.
    }
    \label{fig:R0}
\end{figure}
\begin{figure}[!t]
    \centering
    \includegraphics[width=1.0\linewidth]{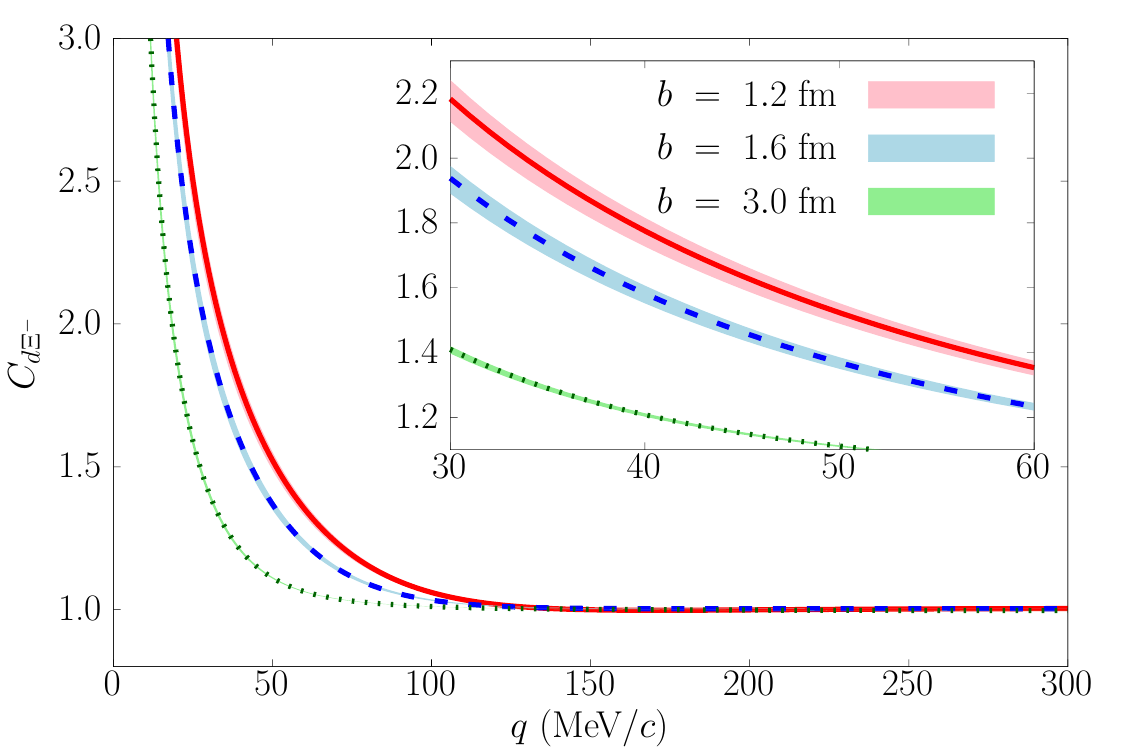}
    \caption{Effect of the source size $b$ on the correlation function. 
    The lines correspond to the results without the 3BF, while the bands of each line are obtained by varying the LECs $G'$ and $H'$ of the 3BF, as in Fig.~\ref{fig:CF}.}
    \label{fig:sourcesize}
\end{figure}
\begin{figure*}[!t]
  \centering
  \includegraphics[width=1.0\linewidth]{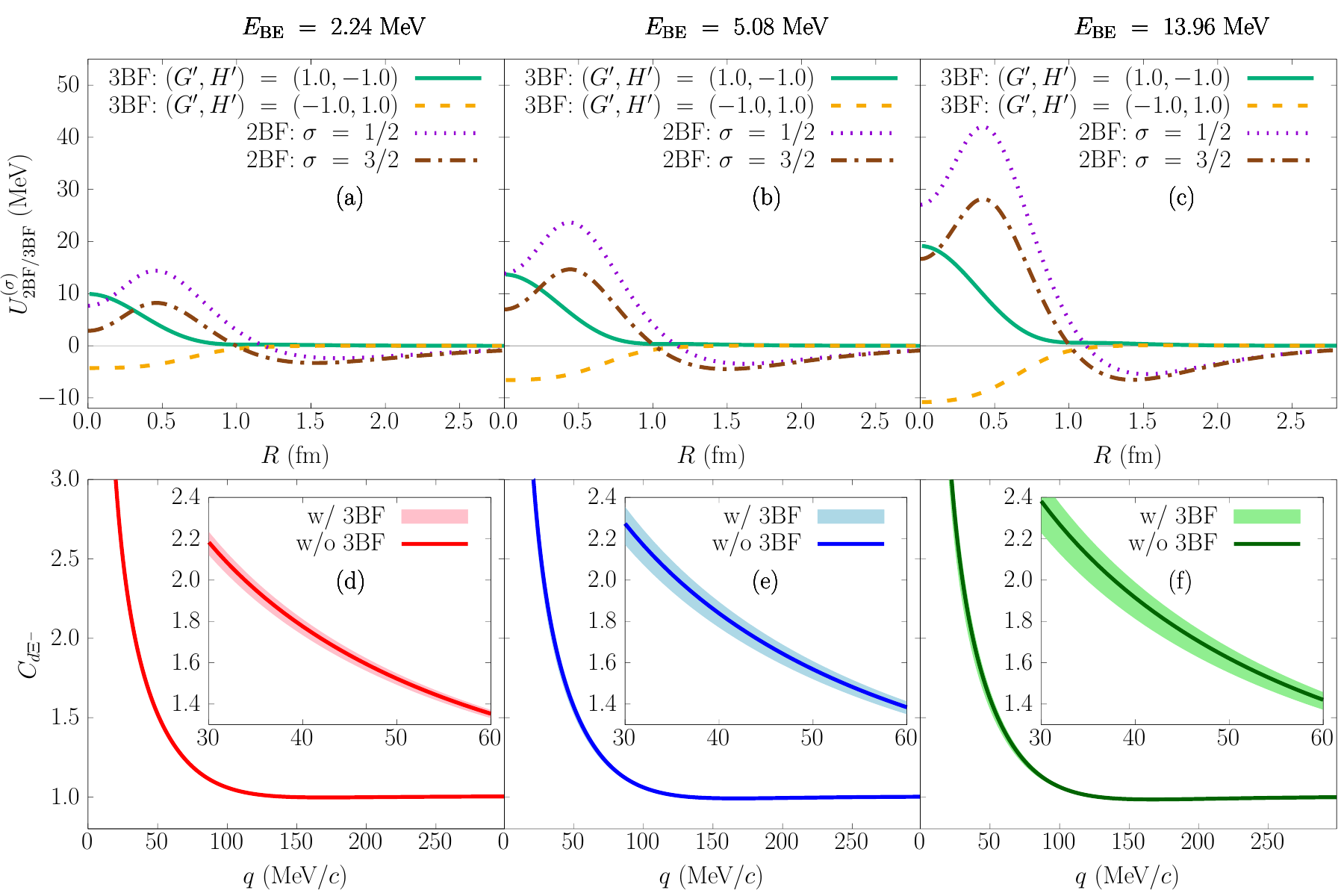}
  \caption{(a)--(c) Behavior of $U_{\mathrm{2BF}}^{(\sigma)}$ and $U_{\mathrm{3BF}}^{(\sigma)}$ with artificially changing the deuteron binding energy $E_{\mathrm{BE}}$ indicated at the top of the figure.
  The labeling of each line follows that of Fig.~\ref{fig:pot}(b).
  (d)--(f) Correlation functions computed with the corresponding $E_{\mathrm{BE}}$. The cutoff and source size are $R_0=1.0$~fm and $b=1.2$~fm, respectively.}
  \label{fig:BE}
\end{figure*}
Here we examine the dependence of $U_{\mathrm{3BF}}^{(\sigma)}$ and $C_{d\Xi^-}$ on the cutoff $R_0$ and the source size $b$.
Figure~\ref{fig:R0}(a) illustrates the $R$ dependence of $U_{\mathrm{3BF}}^{(\sigma)}$ for a 10\% variation of the cutoff around $R_0=1.0$~fm.
The smaller $R_0$ becomes, the greater the absolute value of $U_{\mathrm{3BF}}^{(\sigma)}$, while the range remains almost unchanged. 
Here, we set $G^\prime=1$, $H^\prime=1$, and $b=1.2$~fm. 
The $R_0$ dependence of $C_{d\Xi^-}$ is shown in Fig.~\ref{fig:R0}(b),
where each line is the result calculated by employing the corresponding potential in Fig.~\ref{fig:R0}(a).
While $U_{\mathrm{3BF}}^{(\sigma)}$ near the origin are strongly affected by the choice of $R_0$, $C_{d\Xi^-}$ remains almost unchanged.
Indeed, the three lines in Fig.~\ref{fig:R0}(b) nearly overlap, indicating that the effect of the variation in the cutoff is negligibly small, with the relative change at most $0.1\%$.
This is again attributed to the short-range property of the 3BF.

Figure~\ref{fig:sourcesize} shows the effect of varying the source size $b$ on $C_{d\Xi^-}$. 
Based on the previous study~\cite{PhysRevC.103.065205}, we adopt ${b=1.2}$, $1.6$, and $3.0$~fm, respectively corresponding to the solid, dashed, and dotted lines,
which are the results without the 3BF.
We fix the cutoff $R_0=1.0$~fm.
As $b$ increases, the effect of the 3BF becomes smaller indicated by the narrower bands around the lines.
This behavior can be understood from the form of the source function given by Eq.~\eqref{sourcefunc}.
For larger $b$, the source function spreads over $R$ and its absolute value near the origin decreases. 
This results in the reduced sensitivity to $U_{\mathrm{3BF}}^{(\sigma)}$ at small $R$.

To conclude this section, we consider whether the loosely bound nature of the deuteron can explain the small effect of the $\Xi NN$ 3BF on $C_{d\Xi^-}$.
To examine this, we artificially change the deuteron binding energy $E_{\mathrm{BE}}$ by modulating the strength of the Argonne $V4'$ potential.
Figures~\ref{fig:BE}(a)--(c) show $U_{\mathrm{2BF}}^{(\sigma)}$ and $U_{\mathrm{3BF}}^{(\sigma)}$ computed with $E_{\mathrm{BE}}/\mathrm{nucleon}=1.12$,~$2.54$, and~$6.98$~MeV, corresponding to that of the realistic deuteron, triton and $\alpha$ particle, respectively.
Even though $E_{\mathrm{BE}}$ is increased, the contribution of $U_{\mathrm{3BF}}^{(\sigma)}$ relative to $U_{\mathrm{2BF}}^{(\sigma)}$ does not change significantly.
Figures~\ref{fig:BE}(d)--(f) display $C_{d\Xi^-}$ computed with the potentials in Figs.~\ref{fig:BE}(a)--(c), respectively.
With increasing $E_{\mathrm{BE}}$, the magnitudes of the potentials grow, enhancing $C_{d\Xi^-}$. 
This enhancement mainly originates from $U_{\mathrm{2BF}}^{(\sigma)}$ rather than $U_{\mathrm{3BF}}^{(\sigma)}$.
Indeed, even for the strongly bound case in Fig.~\ref{fig:BE}(f), the effect of the $\Xi NN$ 3BF remains below $8\%$.
Therefore, we conclude that the weak binding of the deuteron is not the primary reason for the small effect of the 3BF on the correlation function.

\section{Summary}
\label{sec:summary}
In this study, we derived for the first time the $\Xi NN$ 3BF based on the SU(3) chiral EFT. 
The LECs of the $\Xi NN$ 3BF were constrained by means of the DSA.
To implement the $\Xi NN$ 3BF in the calculation of correlation functions, we obtained the potential between the deuteron and $\Xi^-$ in coordinate space. 

We evaluated the difficulty of identifying the $\Xi NN$ 3BF from the correlation functions.
Namely, the effect of the $\Xi NN$ 3BF on the deuteron--$\Xi^-$ correlation function was found to be very small, at most around $4\%$. 
Moreover, we found that varying the potential cutoff, source size, or deuteron-binding energy does not significantly change the impact of the $\Xi NN$ 3BF on the correlation function. 
This is because the range of the $\Xi NN$ 3BF is short, in contrast with the longer-range 2BFs. 
In particular, since the contact 3BF potentially gives the dominant contribution, it is difficult to probe the effect of the $\Xi NN$ 3BF through the correlation function, which is primarily sensitive to the potential at medium to long ranges. 


\begin{acknowledgments}
The authors thank T. Hyodo, A. Jinno, S. Ogawa, and Y. Chazono for valuable comments and discussions.
This work was supported by JST ERATO Grant No. JPMJER2304,
as well as JSPS KAKENHI Grants No. JP21K13919 and No. JP23KK0250.
\end{acknowledgments}

\appendix
\begin{widetext}
\section{\uppercase{SU(3) coefficients of vertices relevant to $\Xi NN$ three-baryon force}}
\label{sec:vertexcoef}
Here we list the SU(3) coefficients required for deriving the $\Xi NN$ potentials.
Note that the coefficients corresponding to unlisted combinations of baryons $B$ and mesons $\phi$ are zero.

In Table~\ref{tab:NBBphi} we report the baryon--baryon--meson coefficients $N_{BB\phi}$ relevant to the TPE and OPE potentials.
Table~\ref{tab:NphiBBphi} summarizes the meson--baryon coefficients $N_{\phi\substack{B\\B}\phi}^f$, which enter the TPE potential.
As regards the four-baryon plus one meson coefficients $N_{\substack{BB\\BB}\phi}^f$,
the number of combinations becomes enormous.
Therefore, we list in Table~\ref{tab:NBBBBphisum} the sums over $f$ of the coefficients multiplied by the LECs $D_f$ as in Eqs.~\eqref{N1_1meson} and~\eqref{N2_1meson}.
Similarly, Table~\ref{tab:NBBBBBBsum} summarizes the six-baryon coefficients in a form $\sum_{f}\tilde{t}^{f,a} N_{\substack{BBB\\BBB}}^{f,a}$ as in Eq.~\eqref{VD}.

\begin{table}[!h]
 \begin{center}
  \caption{SU(3) coefficients $N_{BB\phi}$ relevant to the $\Xi NN$-TPE and OPE potentials.
  The axial vector coupling coefficients $g_A$ and $g_B$ are defined in Sec.~\ref{sec:form_XiNNpot_TPE}.}
  \begingroup
  \renewcommand{\arraystretch}{1.4}
  \begin{tabularx}{\columnwidth}{YYYYYYYY}
   \toprule
        $N_{nn\pi^0}$ & $N_{pp\pi^0}$ & $N_{np\pi^-}$ & $N_{pn\pi^+}$ & 
        $N_{\Xi^-\Xi^-\pi^0}$ & $N_{\Xi^0\Xi^0\pi^0}$ & $N_{\Xi^-\Xi^0\pi^-}$ & $N_{\Xi^0\Xi^-\pi^+}$
        \\ \cmidrule(lr){1-8}
        $-g_A$ & $g_A$  & $\sqrt{2}g_A$ & $\sqrt{2}g_A$ & 
        $-g_B$ & $g_B$ & $-\sqrt{2}g_B$ & $-\sqrt{2}g_B$
        \\
   \bottomrule
  \end{tabularx}
  \label{tab:NBBphi}
  \endgroup
 \end{center}
\end{table}

\begin{table}[!t]
 \begin{center}
  \caption{SU(3) coefficients $N_{\phi\substack{B\\B}\phi}^f$ associated with the meson--baryon scattering. The index $f$ specifies the LECs $c_f$ in Eqs.~\eqref{N1_2meson}--\eqref{N3_2meson}, i.e., $c_f = b_D$, $b_F$, $b_0$, $b_1$, $b_2$, $b_3$, $b_4$, $d_1$, and $d_2$.
  When $c_f = b_D$, $b_F$ and $b_0$, the SU(3) coefficients are all expressed by the average pion mass $m_\pi$.}
  \begingroup
  \renewcommand{\arraystretch}{1.4}
  \begin{tabularx}{\columnwidth}{YYYYYYY}
   \toprule
    $c_f$ & 
    $N_{\pi^0\substack{n\\n}\pi^0}^{f}$ &
    $N_{\pi^0\substack{p\\p}\pi^0}^{f}$ &
    $N_{\pi^-\substack{n\\n}\pi^+}^{f}$ &
    $N_{\pi^+\substack{n\\n}\pi^-}^{f}$ &
    $N_{\pi^-\substack{p\\p}\pi^+}^{f}$ &
    $N_{\pi^+\substack{p\\p}\pi^-}^{f}$
    \\ \cmidrule(lr){1-7}
    $b_D$ & 
    $m_\pi^2$ &
    $m_\pi^2$ &
    $m_\pi^2$ &
    $m_\pi^2$ &
    $m_\pi^2$ &
    $m_\pi^2$
    \\
    $b_F$ & 
    $m_\pi^2$ &
    $m_\pi^2$ &
    $m_\pi^2$ &
    $m_\pi^2$ &
    $m_\pi^2$ &
    $m_\pi^2$
    \\
    $b_0$ & 
    $2m_\pi^2$ &
    $2m_\pi^2$ &
    $2m_\pi^2$ &
    $2m_\pi^2$ &
    $2m_\pi^2$ &
    $2m_\pi^2$
    \\ \addlinespace \addlinespace
    $c_f$ & 
    $N_{\pi^0\substack{\Xi^-\\\Xi^-}\pi^0}^{f}$ &
    $N_{\pi^0\substack{\Xi^0\\\Xi^0}\pi^0}^{f}$ &
    $N_{\pi^-\substack{\Xi^-\\\Xi^-}\pi^+}^{f}$ &
    $N_{\pi^+\substack{\Xi^-\\\Xi^-}\pi^-}^{f}$ &
    $N_{\pi^-\substack{\Xi^0\\\Xi^0}\pi^+}^{f}$ &
    $N_{\pi^+\substack{\Xi^0\\\Xi^0}\pi^-}^{f}$
    \\ \cmidrule(lr){1-7}
    $b_D$ & 
    $m_\pi^2$ &
    $m_\pi^2$ &
    $m_\pi^2$ &
    $m_\pi^2$ &
    $m_\pi^2$ &
    $m_\pi^2$
    \\
    $b_F$ & 
    $-m_\pi^2$ &
    $-m_\pi^2$ &
    $-m_\pi^2$ &
    $-m_\pi^2$ &
    $-m_\pi^2$ &
    $-m_\pi^2$
    \\
    $b_0$ & 
    $2m_\pi^2$ &
    $2m_\pi^2$ &
    $2m_\pi^2$ &
    $2m_\pi^2$ &
    $2m_\pi^2$ &
    $2m_\pi^2$
    \\
   \bottomrule
  \end{tabularx}
  \begin{tabularx}{\columnwidth}{YYYYYYYYYYY}
    $c_f$ & 
    $N_{\pi^0\substack{n\\n}\pi^0}^{f}$ & 
    $N_{\pi^0\substack{p\\p}\pi^0}^{f}$ & 
    $N_{\pi^-\substack{n\\n}\pi^+}^{f}$ & 
    $N_{\pi^+\substack{n\\n}\pi^-}^{f}$ & 
    $N_{\pi^-\substack{p\\p}\pi^+}^{f}$ & 
    $N_{\pi^+\substack{p\\p}\pi^-}^{f}$ & 
    $N_{\pi^0\substack{n\\p}\pi^-}^{f}$ & 
    $N_{\pi^-\substack{n\\p}\pi^0}^{f}$ & 
    $N_{\pi^0\substack{p\\n}\pi^+}^{f}$ & 
    $N_{\pi^+\substack{p\\n}\pi^0}^{f}$
    \\ \cmidrule(lr){1-11}
    $b_1$ &
    $1$ &
    $1$ &
    $0$ &
    $2$ &
    $2$ &
    $0$ &
    $\sqrt{2}$ &
    $-\sqrt{2}$ &
    $-\sqrt{2}$ &
    $\sqrt{2}$
    \\
    $b_2$ &
    $1$ &
    $1$ &
    $0$ &
    $2$ &
    $2$ &
    $0$ &
    $\sqrt{2}$ &
    $-\sqrt{2}$ &
    $-\sqrt{2}$ &
    $\sqrt{2}$
    \\
    $b_3$ &
    $1$ &
    $1$ &
    $0$ &
    $2$ &
    $2$ &
    $0$ &
    $\sqrt{2}$ &
    $-\sqrt{2}$ &
    $-\sqrt{2}$ &
    $\sqrt{2}$
    \\
    $b_4$ &
    $2$ &
    $2$ &
    $2$ &
    $2$ &
    $2$ &
    $2$ &
    $0$ &
    $0$ &
    $0$ &
    $0$
    \\
    $d_1$ &
    $0$ &
    $0$ &
    $2$ &
    $-2$ &
    $-2$ &
    $2$ &
    $-2\sqrt{2}$ &
    $2\sqrt{2}$ &
    $2\sqrt{2}$ &
    $-2\sqrt{2}$
    \\
    $d_2$ &
    $0$ &
    $0$ &
    $2$ &
    $-2$ &
    $-2$ &
    $2$ &
    $-2\sqrt{2}$ &
    $2\sqrt{2}$ &
    $2\sqrt{2}$ &
    $-2\sqrt{2}$
    \\ \addlinespace \addlinespace
    $c_f$ & 
    $N_{\pi^0\substack{\Xi^-\\\Xi^-}\pi^0}^{f}$ & 
    $N_{\pi^0\substack{\Xi^0\\\Xi^0}\pi^0}^{f}$ & 
    $N_{\pi^-\substack{\Xi^-\\\Xi^-}\pi^+}^{f}$ & 
    $N_{\pi^+\substack{\Xi^-\\\Xi^-}\pi^-}^{f}$ & 
    $N_{\pi^-\substack{\Xi^0\\\Xi^0}\pi^+}^{f}$ & 
    $N_{\pi^+\substack{\Xi^0\\\Xi^0}\pi^-}^{f}$ & 
    $N_{\pi^0\substack{\Xi^-\\\Xi^0}\pi^-}^{f}$ & 
    $N_{\pi^-\substack{\Xi^-\\\Xi^0}\pi^0}^{f}$ & 
    $N_{\pi^0\substack{\Xi^0\\\Xi^-}\pi^+}^{f}$ & 
    $N_{\pi^+\substack{\Xi^0\\\Xi^-}\pi^0}^{f}$ 
    \\ \cmidrule(lr){1-11}
    $b_1$ &
    $1$ &
    $1$ &
    $0$ &
    $2$ &
    $2$ &
    $0$ &
    $-\sqrt{2}$ &
    $\sqrt{2}$ &
    $\sqrt{2}$ &
    $-\sqrt{2}$
    \\
    $b_2$ &
    $1$ &
    $1$ &
    $0$ &
    $2$ &
    $2$ &
    $0$ &
    $-\sqrt{2}$ &
    $\sqrt{2}$ &
    $\sqrt{2}$ &
    $-\sqrt{2}$
    \\
    $b_3$ &
    $-1$ &
    $-1$ &
    $0$ &
    $-2$ &
    $-2$ &
    $0$ &
    $\sqrt{2}$ &
    $-\sqrt{2}$ &
    $-\sqrt{2}$ &
    $\sqrt{2}$
    \\
    $b_4$ &
    $2$ &
    $2$ &
    $2$ &
    $2$ &
    $2$ &
    $2$ &
    $0$ &
    $0$ &
    $0$ &
    $0$
    \\
    $d_1$ &
    $0$ &
    $0$ &
    $-2$ &
    $2$ &
    $2$ &
    $-2$ &
    $-2\sqrt{2}$ &
    $2\sqrt{2}$ &
    $2\sqrt{2}$ &
    $-2\sqrt{2}$
    \\
    $d_2$ &
    $0$ &
    $0$ &
    $2$ &
    $-2$ &
    $-2$ &
    $2$ &
    $2\sqrt{2}$ &
    $-2\sqrt{2}$ &
    $-2\sqrt{2}$ &
    $2\sqrt{2}$
    \\
   \bottomrule
  \end{tabularx}
    \label{tab:NphiBBphi}
  \endgroup
 \end{center}
\end{table}

\begin{table}[!t]
 \begin{center}
  \caption{Lists of coefficients $\mathcal{M}_{\substack{BB\\BB}\phi}$ and $\mathcal{N}_{\substack{BB\\BB}\phi}$ respectively defined by ${\mathcal{M}_{\substack{BB\\BB}\phi}=\sum_{f=1}^{10} D_f N_{\substack{BB\\BB}\phi}^f}$ and ${\mathcal{N}_{\substack{BB\\BB}\phi}=\sum_{f=11}^{14} D_f N_{\substack{BB\\BB}\phi}^f}$, where the LECs $D_f$ and SU(3) coefficients $N_{\substack{BB\\BB}\phi}^f$ enter the $\Xi NN$-OPE potentials as in Eqs~\eqref{N1_1meson} and~\eqref{N2_1meson}.}
  \begingroup
  \renewcommand{\arraystretch}{1.4}
  \begin{tabularx}{\columnwidth}{XX}
   \toprule
        $\mathcal{M}_{\substack{nn\\nn}\pi^0} = -\mathcal{M}_{\substack{pp\\pp}\pi^0} = -D_1-D_3-D_8-D_{10}$ & 
        $\mathcal{M}_{\substack{np\\np}\pi^0} = -\mathcal{M}_{\substack{pn\\pn}\pi^0} = -D_1+D_3-D_8+D_{10}$
        \\
        $\mathcal{M}_{\substack{np\\nn}\pi^+} = \mathcal{M}_{\substack{nn\\np}\pi^-} = \mathcal{M}_{\substack{pn\\pp}\pi^-} = \mathcal{M}_{\substack{pp\\pn}\pi^+} = \sqrt{2}(D_3+D_{10})$ &
        $\mathcal{M}_{\substack{pn\\nn}\pi^+} = \mathcal{M}_{\substack{nn\\pn}\pi^-} = \mathcal{M}_{\substack{pn\\pp}\pi^-} = \mathcal{M}_{\substack{pp\\np}\pi^+} = \sqrt{2}(D_1+D_8)$
        \\
        $\mathcal{M}_{\substack{n\Xi^-\\n\Xi^-}\pi^0} = \mathcal{M}_{\substack{n\Xi^0\\n\Xi^0}\pi^0} = -\mathcal{M}_{\substack{p\Xi^0\\p\Xi^0}\pi^0} = -\mathcal{M}_{\substack{p\Xi^-\\p\Xi^-}\pi^0} = -D_8$ &
        $\mathcal{M}_{\substack{n\Xi^-\\p\Xi^-}\pi^-} = \mathcal{M}_{\substack{p\Xi^-\\n\Xi^-}\pi^+} = \mathcal{M}_{\substack{p\Xi^0\\n\Xi^0}\pi^+} = \mathcal{M}_{\substack{n\Xi^0\\p\Xi^0}\pi^-} = \sqrt{2}D_8$
        \\
        $\mathcal{M}_{\substack{\Xi^-n\\\Xi^-n}\pi^0} = -\mathcal{M}_{\substack{\Xi^0p\\\Xi^0p}\pi^0} = D_9-D_{10}$ &
        $\mathcal{M}_{\substack{\Xi^0n\\\Xi^0n}\pi^0} = -\mathcal{M}_{\substack{\Xi^-p\\\Xi^-p}\pi^0} = -D_9-D_{10}$
        \\
        $\mathcal{M}_{\substack{n\Xi^-\\\Xi^-n}\pi^0} = -\mathcal{M}_{\substack{p\Xi^0\\\Xi^0p}\pi^0} = \mathcal{M}_{\substack{\Xi^-n\\n\Xi^-}\pi^0} = -\mathcal{M}_{\substack{\Xi^0p\\p\Xi^0}\pi^0} = -D_5+D_7$ &
        $\mathcal{M}_{\substack{p\Xi^-\\\Xi^0n}\pi^0} = -\mathcal{M}_{\substack{n\Xi^0\\\Xi^-p}\pi^0} = -\mathcal{M}_{\substack{\Xi^-p\\n\Xi^0}\pi^0} = \mathcal{M}_{\substack{\Xi^0n\\p\Xi^-}\pi^0} = D_4-D_6$
        \\
        $\mathcal{M}_{\substack{\Xi^-n\\\Xi^-p}\pi^-} = \mathcal{M}_{\substack{\Xi^-p\\\Xi^-n}\pi^+} = \mathcal{M}_{\substack{\Xi^0p\\\Xi^0n}\pi^+} = \mathcal{M}_{\substack{\Xi^0n\\\Xi^0p}\pi^-} = \sqrt{2}D_{10}$ &
        $\mathcal{M}_{\substack{\Xi^-n\\\Xi^0n}\pi^-} = \mathcal{M}_{\substack{\Xi^0n\\\Xi^-n}\pi^+} = \mathcal{M}_{\substack{\Xi^0p\\\Xi^-p}\pi^+} = \mathcal{M}_{\substack{\Xi^-p\\\Xi^0p}\pi^-} = \sqrt{2}D_9$
        \\
        $\mathcal{M}_{\substack{n\Xi^-\\\Xi^-p}\pi^-} = \mathcal{M}_{\substack{p\Xi^0\\\Xi^0n}\pi^+} = \mathcal{M}_{\substack{\Xi^-p\\n\Xi^-}\pi^+} = \mathcal{M}_{\substack{\Xi^0n\\p\Xi^0}\pi^-} = \sqrt{2}(D_4+D_5)$ &
        $\mathcal{M}_{\substack{n\Xi^0\\\Xi^-n}\pi^+} = \mathcal{M}_{\substack{p\Xi^-\\\Xi^0p}\pi^-} = \mathcal{M}_{\substack{\Xi^-n\\n\Xi^0}\pi^-} = \mathcal{M}_{\substack{\Xi^0p\\p\Xi^-}\pi^+} = \sqrt{2}(D_6+D_7)$
        \\
        $\mathcal{M}_{\substack{n\Xi^-\\\Xi^0n}\pi^-} = \mathcal{M}_{\substack{p\Xi^0\\\Xi^-p}\pi^+} = \mathcal{M}_{\substack{\Xi^0n\\n\Xi^-}\pi^+} = \mathcal{M}_{\substack{\Xi^-p\\p\Xi^0}\pi^-} = \sqrt{2}(D_{4}+D_{7})$ &
        $\mathcal{M}_{\substack{p\Xi^-\\\Xi^-n}\pi^+} = \mathcal{M}_{\substack{n\Xi^0\\\Xi^0p}\pi^-} = \mathcal{M}_{\substack{\Xi^-n\\p\Xi^-}\pi^-} = \mathcal{M}_{\substack{\Xi^0p\\n\Xi^0}\pi^+} = \sqrt{2}(D_5+D_6)$
        \\
        \multicolumn{2}{c}{$\mathcal{M}_{\substack{\Xi^0n\\n\Xi^0}\pi^0} = -\mathcal{M}_{\substack{\Xi^-p\\p\Xi^-}\pi^0} = \mathcal{M}_{\substack{n\Xi^0\\\Xi^0n}\pi^0} = -\mathcal{M}_{\substack{p\Xi^-\\\Xi^-p}\pi^0} = -D_4-D_5-D_6-D_7$}
        \\ \cmidrule(lr){1-2}
        $\mathcal{N}_{\substack{\Xi^0n\\n\Xi^0}\pi^0} = -\mathcal{N}_{\substack{\Xi^-p\\p\Xi^-}\pi^0} = D_{12}-D_{13}-D_{14}$ &
        $\mathcal{N}_{\substack{\Xi^-n\\n\Xi^-}\pi^0} = -\mathcal{N}_{\substack{\Xi^0p\\p\Xi^0}\pi^0} = -D_{13}+D_{14}$
        \\
        $\mathcal{N}_{\substack{\Xi^-n\\p\Xi^-}\pi^-} = \mathcal{N}_{\substack{\Xi^0p\\n\Xi^0}\pi^+} = \sqrt{2}D_{13}$ &
        $\mathcal{N}_{\substack{\Xi^0n\\n\Xi^-}\pi^+} = \mathcal{N}_{\substack{\Xi^-p\\p\Xi^0}\pi^-} = \sqrt{2}(-D_{12}+D_{14})$
        \\
        $\mathcal{N}_{\substack{\Xi^-n\\n\Xi^0}\pi^-} = \mathcal{N}_{\substack{\Xi^0p\\p\Xi^-}\pi^+} = \sqrt{2}D_{14}$ &
        $\mathcal{N}_{\substack{\Xi^-p\\n\Xi^-}\pi^+} = \mathcal{N}_{\substack{\Xi^0n\\p\Xi^0}\pi^-} = \sqrt{2}(-D_{12}+D_{13})$
        \\
        \multicolumn{2}{c}{$\mathcal{N}_{\substack{n\Xi^0\\\Xi^0n}\pi^0} = -\mathcal{N}_{\substack{p\Xi^-\\\Xi^-p}\pi^0} = -\mathcal{N}_{\substack{p\Xi^-\\\Xi^0n}\pi^0} = \mathcal{N}_{\substack{n\Xi^0\\\Xi^-p}\pi^0} = -\mathcal{N}_{\substack{\Xi^-p\\n\Xi^0}\pi^0} = \mathcal{N}_{\substack{\Xi^0n\\p\Xi^-}\pi^0} = -D_{12}$}
        \\
        \multicolumn{2}{c}
        {$\mathcal{N}_{\substack{n\Xi^-\\\Xi^-p}\pi^-} = \mathcal{N}_{\substack{p\Xi^0\\\Xi^0n}\pi^+} = \mathcal{N}_{\substack{n\Xi^-\\\Xi^0n}\pi^-} = \mathcal{N}_{\substack{p\Xi^0\\\Xi^-p}\pi^+} = \sqrt{2}D_{12}$}
        \\
   \bottomrule
  \end{tabularx}
    \label{tab:NBBBBphisum}
  \endgroup
 \end{center}
\end{table}

\begin{table}[!t]
 \begin{center}
  \caption{Lists of the contact coefficients $N_{\substack{BBB\\BBB}}^a$ defined by $N_{\substack{BBB\\BBB}}^a=\sum_{f=1}^{11}\tilde{t}^{f,a}
    N_{\substack{BBB\\BBB}}^{f,a}$, which enter the $\Xi NN$-contact potentials as in Eqs.~\eqref{VD}.
    The coefficients are all written in terms of the LECs $C_i$ introduced in Sec.~\ref{sec:form_XiNNpot_ct}.}
  \begingroup
  \renewcommand{\arraystretch}{1.4}
    \footnotesize{
  \begin{tabularx}{\columnwidth}{YYYYYYYYYYYYY}
   \toprule
        $a$ &
        $N_{\substack{n\Xi^-n\\\Xi^-nn}}^a$ & 
        $N_{\substack{p\Xi^0p\\\Xi^0pp}}^a$ &
        $N_{\substack{nn\Xi^-\\\Xi^-nn}}^a$ &
        $N_{\substack{pp\Xi^0\\\Xi^0pp}}^a$ &
        $N_{\substack{\Xi^-nn\\n\Xi^-n}}^a$ &
        $N_{\substack{\Xi^0pp\\p\Xi^0p}}^a$ &
        $N_{\substack{\Xi^-nn\\nn\Xi^-}}^a$ &
        $N_{\substack{\Xi^0pp\\pp\Xi^0}}^a$ &
        $N_{\substack{n\Xi^-p\\\Xi^-np}}^a$ &
        $N_{\substack{p\Xi^0n\\\Xi^0pn}}^a$ &
        $N_{\substack{\Xi^-np\\n\Xi^-p}}^a$ &
        $N_{\substack{\Xi^0pn\\p\Xi^0n}}^a$
        \\ \cmidrule(lr){1-13}
        $1$ & 0 & 0 & 0 & 0 & $C_5$ & $C_5$ &
        $C_2$ & $C_2$ & 0 & 0 & $C_5$ & $C_5$
        \\
        $2$ & $C_{10}$ & $C_{10}$ & 
        $C_7$ & $C_7$ & $C_{10}$ & $C_{10}$ &
        $C_7$ & $C_7$ & $C_{10}$ & $C_{10}$ &
        $C_{10}$ & $C_{10}$
        \\
        $3$ & 0 & 0 & 0 & 0 & 0 & 0 &
        $C_{12}$ & $C_{12}$ & 0 & 0 & 0 & 0
        \\
        $4$ & 0 & 0 & 0 & 0 & $C_{14}$ & $C_{14}$ &
        0 & 0 & 0 & 0 & $C_{14}$ & $C_{14}$
        \\
        $5$ & 0 & 0 & 0 & 0 & 0 & 0 &
        $C_{16}$ & $C_{16}$ & 0 & 0 & 0 & 0
        \\ \addlinespace \addlinespace
        $a$ & 
        $N_{\substack{np\Xi^-\\\Xi^-np}}^a$ &
        $N_{\substack{pn\Xi^0\\\Xi^0pn}}^a$ &
        $N_{\substack{n\Xi^0n\\n\Xi^-p}}^a$ &
        $N_{\substack{p\Xi^-p\\p\Xi^0n}}^a$ &
        $N_{\substack{p\Xi^-n\\n\Xi^-p}}^a$ &
        $N_{\substack{n\Xi^0p\\p\Xi^0n}}^a$ &
        $N_{\substack{\Xi^0nn\\nn\Xi^0}}^a$ &
        $N_{\substack{\Xi^-pp\\pp\Xi^-}}^a$ &
        $N_{\substack{n\Xi^-p\\n\Xi^0n}}^a$ &
        $N_{\substack{p\Xi^0n\\p\Xi^-p}}^a$ &
        $N_{\substack{\Xi^-np\\np\Xi^-}}^a$ &
        $N_{\substack{\Xi^0pn\\pn\Xi^0}}^a$
        \\ \cmidrule(lr){1-13}
        $1$ & 0 & 0 & $C_1$ & $C_1$ & $C_1$ & $C_1$ &
        $C_2$ & $C_2$ & 0 & 0 & $C_2$ & $C_2$
        \\
        $2$ & $C_7$ & $C_7$ & $C_6$ & $C_6$ & $C_6$ & $C_6$ &
        $C_7$ & $C_7$ & $C_6$ & $C_6$ & $C_7$ & $C_7$
        \\
        $3$ & 0 & 0 & $C_{11}$ & $C_{11}$ & $C_{11}$ & $C_{11}$ &
        $C_{12}$ & $C_{12}$ & 0 & 0 & $C_{12}$ & $C_{12}$
        \\
        $4$ & 0 & 0 & 0 & 0 & 0 & 0 & 0 & 0 & 0 & 0 & 0 & 0
        \\
        $5$ & 0 & 0 & $C_{15}$ & $C_{15}$ & $C_{15}$ & $C_{15}$ &
        $C_{16}$ & $C_{16}$ & 0 & 0 & $C_{16}$ & $C_{16}$
        \\ \addlinespace \addlinespace
        $a$ &
        $N_{\substack{n\Xi^0n\\\Xi^-pn}}^a$ &
        $N_{\substack{p\Xi^-p\\\Xi^0np}}^a$ &
        $N_{\substack{\Xi^-pn\\n\Xi^0n}}^a$ &
        $N_{\substack{\Xi^0np\\p\Xi^-p}}^a$ &
        $N_{\substack{n\Xi^0n\\n\Xi^0n}}^a$ &
        $N_{\substack{p\Xi^-p\\p\Xi^-p}}^a$ &
        $N_{\substack{p\Xi^-n\\\Xi^-pn}}^a$ &
        $N_{\substack{n\Xi^0p\\\Xi^0np}}^a$ &
        $N_{\substack{pn\Xi^-\\\Xi^-pn}}^a$ &
        $N_{\substack{np\Xi^0\\\Xi^0np}}^a$ &
        $N_{\substack{\Xi^0nn\\n\Xi^0n}}^a$ &
        $N_{\substack{\Xi^-pp\\p\Xi^-p}}^a$
        \\ \cmidrule(lr){1-13}
        $1$ & $C_3+C_5$ & $C_3+C_5$ & 0 & 0 &
        $C_1$ & $C_1$ & $C_3+C_5$ & $C_3+C_5$ & 0 & 0 &
        $C_5$ & $C_5$
        \\
        $2$ & $C_8+C_{10}$ & $C_8+C_{10}$ &
        $C_8+C_{10}$ & $C_8+C_{10}$ & $2C_6$ & $2C_6$ &
        $C_8+2C_{10}$ & $C_8+2C_{10}$ & $C_7$ & $C_7$ & 
        $C_8+2C_{10}$ & $C_8+2C_{10}$
        \\
        $3$ & 0 & 0 & 0 & 0 & $C_{11}$ & $C_{11}$ & 0 & 0 &
        0 & 0 & 0 & 0
        \\
        $4$ & $C_{13}+C_{14}$ & $C_{13} +C_{14}$ & 
        0 & 0 & 0 & 0 & ${C_{13}+C_{14}}$ & $C_{13}+C_{14}$ & 0 & 0 &
        $C_{14}$ & $C_{14}$
        \\
        $5$ & $C_{17}$ & $C_{17}$ & 0 & 0 & $C_{15}$ & $C_{15}$ &
        $C_{17}$ & $C_{17}$ & 0 & 0 & 0 & 0
        \\ \addlinespace \addlinespace
        $a$ & 
        $N_{\substack{p\Xi^-n\\n\Xi^0n}}^a$ &
        $N_{\substack{n\Xi^0p\\p\Xi^-p}}^a$ &
        $N_{\substack{nn\Xi^0\\\Xi^0nn}}^a$ &
        $N_{\substack{pp\Xi^-\\\Xi^-pp}}^a$ &
        $N_{\substack{n\Xi^-p\\p\Xi^-n}}^a$ &
        $N_{\substack{p\Xi^0n\\n\Xi^0p}}^a$ &
        $N_{\substack{n\Xi^0n\\\Xi^0nn}}^a$ &
        $N_{\substack{p\Xi^-p\\\Xi^-pp}}^a$ &
        $N_{\substack{\Xi^-pn\\p\Xi^-n}}^a$ &
        $N_{\substack{\Xi^0np\\n\Xi^0p}}^a$ &
        $N_{\substack{n\Xi^0n\\p\Xi^-n}}^a$ &
        $N_{\substack{p\Xi^-p\\n\Xi^0p}}^a$
        \\ \cmidrule(lr){1-13}
        $1$ & $C_1$ & $C_1$ & 0 & 0 & 0 & 0 &
        $C_3+C_5$ & $C_3+C_5$ & $C_5$ & $C_5$ & 0 & 0
        \\
        $2$ & $C_6$ & $C_6$ & $C_7$ & $C_7$ & $C_6$ & $C_6$ &
        $C_8+2C_{10}$ & $C_8+2C_{10}$ & $C_8+2C_{10}$ & $C_8+2C_{10}$ &
        $C_6$ & $C_6$
        \\
        $3$ & $C_{11}$ & $C_{11}$ & 0 & 0 & 0 & 0 & 0 & 0 & 0 & 0 & 0 & 0
        \\
        $4$ & 0 & 0 & 0 & 0 & 0 & 0 & $C_{13}+C_{14}$ & $C_{13}+C_{14}$ &
        $C_{14}$ & $C_{14}$ & 0 & 0
        \\
        $5$ & $C_{15}$ & $C_{15}$ & 0 & 0 & 0 & 0 & $C_{17}$ & $C_{17}$ &
        0 & 0 & 0 & 0
        \\
  \end{tabularx}
  \begin{tabularx}{\columnwidth}{YYYYYYY}
       \toprule
        $a$ &
        $N_{\substack{\Xi^-pn\\pn\Xi^-}}^a$ &
        $N_{\substack{\Xi^0np\\np\Xi^0}}^a$ &
        $N_{\substack{\Xi^0nn\\p\Xi^-n}}^a$ &
        $N_{\substack{\Xi^-pp\\n\Xi^0p}}^a$ &
        $N_{\substack{p\Xi^-n\\\Xi^0nn}}^a$ &
        $N_{\substack{n\Xi^0p\\\Xi^-pp}}^a$
        \\ \cmidrule(lr){1-7}
        $1$ & $C_2$ & $C_2$ & 0 & 0 & $C_3+C_5$ & $C_3+C_5$
        \\
        $2$ & $C_7$ & $C_7$ & $C_8+C_{10}$ & $C_8+C_{10}$ &
        $C_8+C_{10}$ & $C_8+C_{10}$
        \\
        $3$ & $C_{12}$ & $C_{12}$ & 0 & 0 & 0 & 0
        \\
        $4$ & 0 & 0 & 0 & 0 & $C_{13}+C_{14}$ & $C_{13}+C_{14}$
        \\
        $5$ & $C_{16}$ & $C_{16}$ & 0 & 0 & $C_{17}$ & $C_{17}$
        \\
   \bottomrule
  \end{tabularx}
  }
    \label{tab:NBBBBBBsum}
  \endgroup
 \end{center}
\end{table}

\clearpage
\section{\uppercase{Detail of calculating deuteron--$\Xi^-$ potential in coordinate space}}
\label{sec:dXipot_app}
\subsection{Central components of $\Xi NN$ potential}
\label{sec:dXipot_app_central}
To calculate Eq.~\eqref{WrR}, we first need to derive the central components of the $\Xi NN$ potential $V_{(0)}^{\Xi NN}$ in momentum space.
Here, the subscript $(0)$ represents the rank-0 component of the irreducible tensors that form the 3BF.
By performing the irreducible tensor decomposition~\cite{FUKUI2024138839},
the central components of the $\Xi NN$ potential are given by
\begin{align}
    V_{(0)}^{\Xi NN}(\bm{q}_1,\bm{q}_2,\bm{q}_3)
    &=
    V_{\mathrm{TPE}(0)}^{\Xi NN}(\bm{q}_1,\bm{q}_2,\bm{q}_3)
    +V_{\mathrm{OPE}(0)}^{\Xi NN}(\bm{q}_1,\bm{q}_2,\bm{q}_3)
    +V_{\mathrm{ct}(0)}^{\Xi NN}(\bm{q}_1,\bm{q}_2,\bm{q}_3),
    \label{V0sum} 
\end{align}
where the TPE term reads
\begin{align}
    V_{\mathrm{TPE}(0)}^{\Xi NN}(\bm{q}_1,\bm{q}_2,\bm{q}_3)
    &=
    \mathcal{A}_{23}\left[Y_{123(0)}^{456}+Y_{231(0)}^{564}+Y_{312(0)}^{645}\right],
    \label{VTPE0}
\end{align}
with
\begin{align}
    Y_{123(0)}^{456}
    &=
    \begin{aligned}[t]
    \dfrac{g_Ag_B}{36f_\pi^4}\dfrac{1}{\left(q_{1}^2+m_\pi^2\right)\left(q_{3}^2+m_\pi^2\right)}
    &\left[\bm{\tau}_1\cdot\bm{\tau}_3\left\{4c_1 m_\pi^2 \left(\bm{\sigma}_1\cdot\bm{\sigma}_3\right)\left(\bm{q}_{1}\cdot\bm{q}_{3}\right)
    -2c_3\left(\bm{\sigma}_1\cdot\bm{\sigma}_3\right)\left(\bm{q}_{1}\cdot\bm{q}_{3}\right)^2
\right\}\right.
    \\
    &\left.
    +\frac{c_4}{2}\left\{\bm{\tau}_3\cdot(\bm{\tau}_1\times\bm{\tau}_2)\right\}
    \left\{\left(\bm{\sigma}_1\times\bm{\sigma}_3\right)\cdot\bm{\sigma}_2\right\}
    \left\{q_1^2 q_3^2 -\left(\bm{q}_{1}\cdot\bm{q}_{3}\right)^2\right\}\right],
    \end{aligned}
    \label{Y1234560}\\
    Y_{231(0)}^{564}
    &=
    \begin{aligned}[t]
    \dfrac{g_Ag_B}{36f_\pi^4}\dfrac{1}{\left(q_{2}^2+m_\pi^2\right)\left(q_{1}^2+m_\pi^2\right)}
    &\left[\bm{\tau}_2\cdot\bm{\tau}_1\left\{4c_1 m_\pi^2
    \left(\bm{\sigma}_2\cdot\bm{\sigma}_1\right)\left(\bm{q}_{2}\cdot\bm{q}_{1}\right)
    -2c_3\left(\bm{\sigma}_2\cdot\bm{\sigma}_1\right)\left(\bm{q}_{2}\cdot\bm{q}_{1}\right)^2
\right\}\right.
    \\
    &\left.
    +\frac{c_4}{2}\left\{\bm{\tau}_3\cdot(\bm{\tau}_1\times\bm{\tau}_2)\right\}
    \left\{\left(\bm{\sigma}_2\times\bm{\sigma}_1\right)\cdot\bm{\sigma}_3\right\}
    \left\{q_2^2 q_1^2 -\left(\bm{q}_{2}\cdot\bm{q}_{1}\right)^2\right\}\right],
    \end{aligned}
    \label{Y2315640}\\
    Y_{312(0)}^{645}
    &=
    \begin{aligned}[t]
    \dfrac{g_A^2}{36f_\pi^4}\dfrac{1}{\left(q_{3}^2+m_\pi^2\right)\left(q_{2}^2+m_\pi^2\right)}
    &\left[\bm{\tau}_3\cdot\bm{\tau}_2\left\{-12u_1 m_\pi^2
    \left(\bm{\sigma}_3\cdot\bm{\sigma}_2\right)\left(\bm{q}_{3}\cdot\bm{q}_{2}\right)
    +6u_3\left(\bm{\sigma}_3\cdot\bm{\sigma}_2\right)\left(\bm{q}_{3}\cdot\bm{q}_{2}\right)^2
\right\}\right.
    \\
    &\left.
    -\frac{u_4}{2}\bm{\tau}_3\cdot(\bm{\tau}_1\times\bm{\tau}_2)
    \left\{\left(\bm{\sigma}_3\times\bm{\sigma}_2\right)\cdot\bm{\sigma}_1\right\}
    \left\{q_3^2 q_2^2 -\left(\bm{q}_{3}\cdot\bm{q}_{2}\right)^2\right\}\right].
    \end{aligned}
    \label{Y3126450}
\end{align}
The OPE term is given by
\begin{align}
    V_{\mathrm{OPE}(0)}^{\Xi NN}(\bm{q}_1,\bm{q}_2,\bm{q}_3)
    &=
    X_{123(0)}^{456}+X_{231(0)}^{564}+X_{312(0)}^{645}
    +\mathcal{P}_{23}\mathcal{P}_{12}X_{312(0)}^{564}
    +\mathcal{P}_{23}\mathcal{P}_{13}X_{231(0)}^{645},
    \label{VOPE0}
\end{align}
with
\begin{align}
    X_{123(0)}^{456}
    &=
    \dfrac{g_Bd}{6f_\pi^2}\dfrac{q_{1}^2}{q_{1}^2+m_\pi^2}
    \left[(\bm{\tau}_2-\bm{\tau}_3)\cdot\bm{\tau}_1\,(\bm{\sigma}_2-\bm{\sigma}_3)\cdot\bm{\sigma}_1+(\bm{\tau}_1\times\bm{\tau}_2)\cdot\bm{\tau}_3\,(\bm{\sigma}_1\times\bm{\sigma}_2)\cdot\bm{\sigma}_{3}\right],
    \label{X1234560}\\
    X_{231(0)}^{564}
    &=
    \dfrac{g_A}{6f_\pi^2}\dfrac{q_{2}^2}{q_{2}^2+m_\pi^2}
    \left[(e_1\bm{\sigma}_3+e_2\bm{\sigma}_1)\cdot\bm{\sigma}_2\,\bm{\tau}_2\cdot\bm{\tau}_3
    +(e_3\bm{\sigma}_3+e_4\bm{\sigma}_1)\cdot\bm{\sigma}_2\,\bm{\tau}_1\cdot\bm{\tau}_2-e_5(\bm{\tau}_1\times\bm{\tau}_2)\cdot\bm{\tau}_3\,(\bm{\sigma}_1\times\bm{\sigma}_2)\cdot\bm{\sigma}_3\right],
    \label{X2315640}\\
    X_{312(0)}^{645}
    &=
    \dfrac{g_A}{6f_\pi^2}\dfrac{q_{3}^2}{q_{3}^2+m_\pi^2}
    \left[(e_4\bm{\sigma}_1+e_3\bm{\sigma}_2)\cdot\bm{\sigma}_3\,\bm{\tau}_3\cdot\bm{\tau}_1+(e_2\bm{\sigma}_1+e_1\bm{\sigma}_2)\cdot\bm{\sigma}_3\,\bm{\tau}_2\cdot\bm{\tau}_3-e_5(\bm{\tau}_1\times\bm{\tau}_2)\cdot\bm{\tau}_3\,(\bm{\sigma}_1\times\bm{\sigma}_2)\cdot\bm{\sigma}_3\right],
    \label{X3126450}\\
    X_{312(0)}^{564}
    &=
    \dfrac{g_A}{6f_\pi^2}\dfrac{q_{3}^2}{q_{3}^2+m_\pi^2}
    \left[\{f_1\bm{\sigma}_1+f_2\bm{\sigma}_2+f_3i(\bm{\sigma}_1\times\bm{\sigma}_2)\}\cdot\bm{\sigma}_3\,\bm{\tau}_3\cdot\bm{\tau}_1+\{f_2\bm{\sigma}_1+f_1\bm{\sigma}_2+f_3i(\bm{\sigma}_1\times\bm{\sigma}_2)\}\cdot\bm{\sigma}_3\,\bm{\tau}_2\cdot\bm{\tau}_3\right.
    \notag\\
    &\left.+\{f_4(\bm{\sigma}_1+\bm{\sigma}_2)+if_5(\bm{\sigma}_1\times\bm{\sigma}_2)\}\cdot\bm{\sigma}_3\,i(\bm{\tau}_1\times\bm{\tau}_2)\cdot\bm{\tau}_3\right],
    \label{X3125640}\\
    X_{231(0)}^{645}
    &=
    \dfrac{g_A}{6f_\pi^2}\dfrac{q_{2}^2}{q_{2}^2+m_\pi^2}
    \left[\{f_1\bm{\sigma}_3+f_2\bm{\sigma}_1-f_3i(\bm{\sigma}_3\times\bm{\sigma}_1)\}\cdot\bm{\sigma}_2\,\bm{\tau}_2\cdot\bm{\tau}_3+\{f_2\bm{\sigma}_3+f_1\bm{\sigma}_1-f_3i(\bm{\sigma}_3\times\bm{\sigma}_1)\}\cdot\bm{\sigma}_2\,\bm{\tau}_1\cdot\bm{\tau}_2\right.\notag\\
    &\left.-\{f_4(\bm{\sigma}_3+\bm{\sigma}_1)-if_5(\bm{\sigma}_3\times\bm{\sigma}_1)\}\cdot\bm{\sigma}_2\,i(\bm{\tau}_1\times\bm{\tau}_2)\cdot\bm{\tau}_3\right].
    \label{X2316450}
\end{align}
The contact term includes the central component only;
\begin{align}
    V_{\mathrm{ct}(0)}^{\Xi NN}(\bm{q}_1,\bm{q}_2,\bm{q}_3)
    &=
    V_{\mathrm{ct}}^{\Xi NN},
    \label{Vct0}
\end{align}
where $V_{\mathrm{ct}}^{\Xi NN}$ is defined by Eq.~\eqref{eq:ct}.

\subsection{Matrix elements of spin--isospin operators}
\label{sec:dXipot_app_spinisospin}
Equation~\eqref{WrR} contains the matrix elements of $V_{(0)}^{\Xi NN}$ in the spin--isospin states. 
The spin state $\Ket{\Upsilon_{\sigma m_{\sigma}}}$ and isospin state $\Ket{\Theta_{1/2,-1/2}}$ of the $p$--$n$--$\Xi^-$ three-body system are defined as
\begin{align}
    \Ket{\Upsilon_{\sigma m_{\sigma}}}
    &=
    \Ket{\biggl[\left[\eta_{1/2}^{(p)}\otimes\eta_{1/2}^{(n)}\right]_1\otimes\eta_{1/2}^{(\Xi^-)}\biggr]_{\sigma m_\sigma}},
    \label{Usilon}\\
    \Ket{\Theta_{1/2,-1/2}}
    &=
    \Ket{\biggl[\left[\zeta^{(p)}_{1/2}\otimes\zeta^{(n)}_{1/2}\right]_0\otimes\zeta^{(\Xi^-)}_{1/2}\biggr]_{1/2,-1/2}},
    \label{Theta}
\end{align}
where the spin (isospin) state of particle $\alpha$ is denoted by $\Ket{\eta_{1/2}^{(\alpha)}}$ $\left(\Ket{\zeta_{1/2}^{(\alpha)}}\right)$.
The tensor coupling of the angular momentum is defined by
\begin{align}
    \Ket{\left[\eta_{1/2}^{(p)}\otimes\eta_{1/2}^{(n)}\right]_{1\mu}}
    =
    \sum_{m_p m_n}
    \left(\left. \frac{1}{2} m_p \frac{1}{2} m_n \right| 1 \mu \right)
    \Ket{\eta_{1/2 \, m_p}^{(p)}}
    \Ket{\eta_{1/2 \, m_n}^{(n)}},
    \label{tensor}
\end{align}
with $\mu=\pm 1$.
In Eq.~\eqref{Usilon}, $\sigma$ and $m_\sigma$ represent the total spin of the system and its third component, respectively. 
The isospin and its third component of the system are fixed as $1/2$ and $-1/2$, respectively.
Using these states, we evaluate the matrix elements of all spin and isospin operators involved in the potentials given in Sec.~\ref{sec:dXipot_app_central}. 

Hereafter we redefine the particle indices $\alpha$ as $\Xi\to1$, $p\to2$, and $n\to3$.
The matrix elements of the spin operators, $\bm{\sigma}_1\cdot\bm{\sigma}_2$ and $\bm{\sigma}_1\cdot\bm{\sigma}_3$, depend on $\sigma$:
\begin{align}
    \Braket{\Upsilon_{\sigma m_\sigma}|\bm{\sigma}_1\cdot\bm{\sigma}_2|\Upsilon_{\sigma m_\sigma}}
    &=
    \Braket{\Upsilon_{\sigma m_\sigma}|\bm{\sigma}_3\cdot\bm{\sigma}_1|\Upsilon_{\sigma m_\sigma}}
    =
    \begin{dcases}
    -2 &(\sigma=1/2)\\
    1 &(\sigma=3/2).
    \end{dcases}
    \label{sig1sig2}
\end{align}
The other spin operators yield $\sigma$-independent expressions:
\begin{gather}
    \Braket{\Upsilon_{\sigma m_\sigma}|\bm{\sigma}_2\cdot\bm{\sigma}_3|\Upsilon_{\sigma m_\sigma}}=1,
    \label{sig2sig3}\\
    \Braket{\Upsilon_{\sigma m_\sigma}|\left(\bm{\sigma}_1\times\bm{\sigma}_2\right)\cdot\bm{\sigma}_3|\Upsilon_{\sigma m_\sigma}}=0.
    \label{sig1sig2sig3}
\end{gather}
Similarly, the matrix elements of the isospin operators can be calculated as
\begin{align}
    &\Braket{\Theta_{1/2,-1/2}|\bm{\tau}_1\cdot\bm{\tau}_2|\Theta_{1/2,-1/2}}
    =
    \Braket{\Theta_{1/2,-1/2}|\bm{\tau}_3\cdot\bm{\tau}_1|\Theta_{1/2,-1/2}}
    =0,
    \label{tau1tau2}\\
    &\Braket{\Theta_{1/2,-1/2}|\bm{\tau}_2\cdot\bm{\tau}_3|\Theta_{1/2,-1/2}}=-3,
    \label{tau2tau3}\\
    &\Braket{\Theta_{1/2,-1/2}|\left(\bm{\tau}_1\times\bm{\tau}_2\right)\cdot\bm{\tau}_3|\Theta_{1/2,-1/2}}=0.
    \label{tau1tau2tau3}
\end{align}

\subsection{Fourier transform and multipole expansion}
\label{sec:dXipot_app_FourierMPE}
As shown in the previous section, there are some matrix elements that vanish for the $d$--$\Xi^-$ system, and therefore, only a few terms involved in $V_{(0)}^{\Xi NN}$ remain finite after evaluating the spin--isospin expectation values.
For example, by computing the matrix element of the central TPE potential $V_{\mathrm{TPE}(0)}^{\Xi NN}$, only the terms dependent on the LECs $u_1$ and $u_3$ contribute.

Then, we can analytically calculate the momentum integration in Eq.~\eqref{WrR}.
As a representative example, we consider the TPE-$u_3$ term.
Using the residue theorem, the Fourier transform is performed as
\begin{samepage}
\begin{align}
    &\frac{1}{(2\pi)^6} \iint d\bm{q}_2 d\bm{q}_3
    \exp\left[i(\bm{q}_2\cdot\bm{r}_2 + \bm{q}_3\cdot\bm{r}_3)\right]
    \Braket{\Upsilon_{\sigma m_\sigma}\Theta_{1/2,-1/2}|
    V_{u_3(0)}^{\Xi NN}(\bm{q}_1,\bm{q}_2,\bm{q}_3)
    |\Upsilon_{\sigma m_\sigma}\Theta_{1/2,-1/2}}
    \notag\\
    &\quad=
    \frac{g_A^2 u_3}{6f_\pi^4}
    \Braket{\Theta_{1/2,-1/2} | \bm{\tau}_3\cdot\bm{\tau}_2 | \Theta_{1/2,-1/2}}
    \Braket{\Upsilon_{\sigma m_\sigma}|\bm{\sigma}_3\cdot\bm{\sigma}_2 | \Upsilon_{\sigma m_\sigma}}
    \notag\\
    &\quad\times
    \frac{1}{(2\pi)^6}\iint d\bm{q}_2 d\bm{q}_3
    \exp\left[i(\bm{q}_2\cdot\bm{r}_2 + \bm{q}_3\cdot\bm{r}_3)\right]
    \dfrac{\left(\bm{q}_{3}\cdot\bm{q}_{2}\right)^2}{\left(q_{3}^2+m_\pi^2\right)\left(q_{2}^2+m_\pi^2\right)}
    \notag\\
    &\quad=
    -\frac{g_A^2 u_3}{2f_\pi^4}
    \frac{1}{(2\pi)^6}\nabla_{\bm{r}_2}^2\nabla_{\bm{r}_3}^2
    \left[
    \frac{(4\pi)^2}{3}
    \int dq_2 \frac{q_2^2}{q_2^2+m_\pi^2}j_0(q_2r_2)
    \int dq_3 \frac{q_3^2}{q_3^2+m_\pi^2}j_0(q_3r_3)
    \right.
    \notag\\
    &\qquad\left.
    +\frac{2(4\pi)^3}{3\sqrt{5}}
    \left[Y_2\left(\hat{\bm{r}}_{2}\right)\otimes Y_2\left(\hat{\bm{r}}_{3}\right)\right]_{00}
    \int dq_2 \frac{q_2^2}{q_2^2+m_\pi^2}j_2(q_2r_2)
    \int dq_3 \frac{q_3^2}{q_3^2+m_\pi^2}j_2(q_3r_3)
    \right]
    \notag\\
    &\quad=
    -\frac{g_A^2 u_3 m_\pi^6}{12\pi f_\pi^4}
    \left[
    \frac{1}{8\pi}Y(m_\pi r_2)Y(m_\pi r_3)
    +\frac{1}{\sqrt{5}}
    Z(m_\pi r_2) Z(m_\pi r_3)
    \left[Y_2\left(\hat{\bm{r}}_{2}\right)\otimes Y_2\left(\hat{\bm{r}}_{3}\right)\right]_{00}
    \right],
    \label{WrRu3}
\end{align}
\end{samepage}
where the functions $Y$ and $Z$ are defined by Eqs.~\eqref{Yx} and~\eqref{Zx}, respectively.
The arguments of the spherical harmonics $Y_{\lambda\mu}$ ($-\lambda \leq \mu \leq \lambda$) are defined by $\hat{\bm{r}}_i=\bm{r}_i/r_i$.
To carry out the above calculations, the following relation is used~\cite{Machleidt1986}:
\begin{align}
    q^2\exp(i\bm{q}\cdot\bm{r})=-\nabla_{\bm{r}}^2\exp(i\bm{q}\cdot\bm{r}),
    \label{q2Laplacian}
\end{align}
as well as
\begin{align}
    \left(\hat{\bm{q}}_{3}\cdot\hat{\bm{q}}_{2}\right)^2
    &=
    \frac{1}{3}
    +\frac{8\pi}{3\sqrt{5}}\left[Y_2\left(\hat{\bm{q}}_{3}\right)\otimes Y_2\left(\hat{\bm{q}}_{2}\right)\right]_{00},
    \label{qqYY}\\
    \exp\left(i\bm{q}_2\cdot\bm{r}_2\right)
    &=
    4\pi\sum_\lambda (-i)^\lambda \hat{\lambda}\,
    j_\lambda(q_2r_2)
    \left[Y_\lambda\left(\hat{\bm{q}}_2\right)\otimes Y_\lambda\left(\hat{\bm{r}}_2\right)\right]_{00}.
    \label{Rayleigh}
\end{align}
Here, $j_\lambda$ is the spherical Bessel function.

Consequently, the thus-obtained potential $W^{(\sigma)}$ reads
\begin{align}
    W^{(\sigma)}(\bm{r},\bm{R})
    &=
    W_{\mathrm{TPE}}(\bm{r},\bm{R})
    +W_{\mathrm{OPE}}^{(\sigma)}(\bm{r},\bm{R})
    +W_{\mathrm{ct}}^{(\sigma)}(\bm{r},\bm{R}),
    \label{Wsum}\\    
    W_{\mathrm{TPE}}(\bm{r},\bm{R})
    &=
    -\frac{g_A^2 m_\pi^6 u_1}{4\sqrt{3}\pi f_\pi^4}
    Z_1(m_\pi r_2) Z_1(m_\pi r_3)
    \left[Y_1\left(\hat{\bm{r}}_2\right)\otimes Y_1\left(\hat{\bm{r}}_3\right)\right]_{00}
    \notag\\
    &-\dfrac{g_A^2 m_\pi^6 u_3}{12\pi f_\pi^4}\left[\dfrac{1}{8\pi}
    Y(m_\pi r_2) Y(m_\pi r_3)
    +\dfrac{1}{\sqrt{5}}Z(m_\pi r_2)Z(m_\pi r_3)
    \left[Y_2\left(\hat{\bm{r}}_2\right)\otimes Y_3\left(\hat{\bm{r}}_2\right)\right]_{00}\right],
    \label{WTPE}\\    
    W_{\mathrm{OPE}}^{(\sigma)}(\bm{r},\bm{R})
    &=
    -\dfrac{g_A m_\pi^3}{4\pi f_\pi^2}
    \left[
    \dfrac{G_e^{(\sigma)}}{2} \left[Y(m_\pi r_2)\delta(\bm{r}_3) + Y(m_\pi r_3)\delta(\bm{r}_2)+\right]
    +3G_f^{(\sigma)}\left[Y(m_\pi r_2)\delta(\bm{r}) + Y(m_\pi r_3)\delta(\bm{r})\right]\right],
    \label{WOPE}\\    
    W_{\mathrm{ct}}^{(\sigma)}(\bm{r},\bm{R})
    &=
    H^{(\sigma)}\delta(\bm{r}_2)\delta(\bm{r}_3),
    \label{Wct}    
\end{align}
where $\bm{r}_2$ and $\bm{r}_3$ can be expressed as functions of $\bm{r}$ and $\bm{R}$ by Eqs.~\eqref{r2} and~\eqref{r3}.
The function $Z_1$, as well as the constants $G_e^{(\sigma)}$, $G_f^{(\sigma)}$, and $H^{(\sigma)}$, are all defined in Sec.~\ref{sec:form_dXipot}.

Then, by applying the regularization described in Sec.~\ref{sec:form_dXipot} to $W^{(\sigma)}$, we can perform the integration over $\bm{r}$ and $\hat{\bm{R}}$ in Eq.~\eqref{Ufold_general},
for which the multipole expansion is applied to the functions dependent on $r_2$ and $r_3$.
For example, 
\begin{align}
    Y(m_\pi r_2)
    &=
    4\pi \sum_{\lambda}\frac{(-)^\lambda}{\hat{\lambda}}
    \mathcal{F}_\lambda(r,R)
    \left[Y_\lambda\left(\hat{\bm{r}}\right) \otimes Y_\lambda\left(\hat{\bm{R}}\right)\right]_{00},
    \label{MPE}\\
    \mathcal{F}_\lambda(r,R)
    &=
    \frac{\hat{\lambda}^2}{2}
    \int_{-1}^{1}dw P_\lambda(w) Y\!\left(m_\pi\sqrt{M_2^2 r^2+R^2-2M_2rR w}\right),
    \label{MPEfunc}
\end{align}
where $w=\cos\theta$ and $\theta$ is the angle between $\bm{r}$ and $\bm{R}$.
Accordingly, the spherical harmonics dependent on $\hat{\bm{r}}_2$ is also expressed in terms of $\hat{\bm{r}}$ and $\hat{\bm{R}}$ as
\begin{align}
    r_2^\lambda Y_{\lambda\mu}\left(\hat{\bm{r}}_2\right)
    =
    \sum_{\Lambda=0}^\lambda
    \frac{\sqrt{4\pi}}{\hat\Lambda} \left(\mqty{2\lambda+1 \\ 2\Lambda}\right)^{\frac{1}{2}}
    \left(-M_2 r\right)^{\lambda-\Lambda}\, R^{\Lambda}
    \left[Y_{\lambda-\Lambda}\left(\hat{\bm{r}}\right) \otimes Y_\Lambda\left(\hat{\bm{R}}\right)\right]_{\lambda\mu},
\end{align}
and similar for other spherical harmonics dependent on $\hat{\bm{r}}_3$.
The binomial coefficient $\left(\mqty{2\lambda+1 \\ 2\Lambda}\right)$ is defined by Eq.~\eqref{binomial}.
Thus, finally we derive Eqs.~\eqref{UTPEu1}--\eqref{Uct}.
    
\end{widetext}

\bibliography{XiNN}

\end{document}